\documentclass[draftcls,onecolumn]{IEEEtran}
\usepackage{epsfig,latexsym,amsmath,amssymb,setspace,cite,color,graphicx,cases}

\PassOptionsToPackage{bookmarks=false}{hyperref}

\usepackage[colorlinks=true, linkcolor=blue, citecolor=blue, urlcolor=blue]{hyperref}

\usepackage{physics}

\makeatletter

\newcommand{\pushright}[1]{\ifmeasuring@#1\else\omit\hfill$\displaystyle#1$\fi\ignorespaces}
\makeatother

\usepackage{amsmath,amssymb,amsfonts}
\usepackage{multicol}
\usepackage{amsmath}
\usepackage{mleftright,mathtools}
\usepackage{amsfonts}
\usepackage{amssymb}
\usepackage{dsfont}
\usepackage{bm}
\usepackage{amsthm}
\usepackage{newlfont}
\usepackage{float}
\usepackage{algorithm}
\usepackage{algpseudocode}
\usepackage{enumerate}
\usepackage{chngcntr}
\usepackage{mathtools}
\usepackage{breqn}
\usepackage{stmaryrd}
\usepackage[caption=false,font=footnotesize]{subfig}
\usepackage{cite}
\usepackage{tikz}
\usepackage{stmaryrd}
\usepackage{multirow}
\tikzset{
  person/.style = {pic={man,scale=#1}},   
  person/.default = 0.9,                  
  arrow/.style  = {->,>=Latex,thick},
}
\usetikzlibrary{positioning,arrows.meta,calc}
\usetikzlibrary{shapes.geometric, arrows.meta, positioning}

\hypersetup{
    colorlinks=true,  
    linkcolor=blue,  
    citecolor=blue,  
        urlcolor=blue  
}
\DeclareMathAlphabet\mathbfcal{OMS}{cmsy}{b}{n}

\allowdisplaybreaks

\usepackage{enumitem}

\makeatletter
\newlist{algsub}{enumerate}{1}
\setlist[algsub,1]{
  label=\arabic{ALG@line}.\arabic*,   
  ref=\arabic{ALG@line}.\arabic*,
  leftmargin=*,                      
  nosep                               
}
\makeatother
\newcounter{algsubstate}

\newenvironment{algsubstates}
  {\setcounter{algsubstate}{0}
   \renewcommand{\State}{
     \refstepcounter{algsubstate}
     \Statex {\footnotesize\alph{algsubstate}:}\space}}
  {}

\newtheorem{definition}{Definition}
\newtheorem{lemma}{Lemma}
\newtheorem{example}{Example}

\newtheorem{theorem}{Theorem}
\newtheorem{corollary}{Corollary}
\newtheorem*{properties}{Properties}

\newtheorem{proposition}{Proposition}
\newtheorem{remark}{Remark}
\begin{document}

\sloppy
\renewcommand{\qedsymbol}{ \begin{tiny}$\blacksquare$ \end{tiny} }

\renewcommand{\algorithmicrequire}{\textbf{Input:}}
\renewcommand{\algorithmicensure}{\textbf{Inputs:}}

\renewcommand{\leq}{\leqslant}
\renewcommand{\geq}{\geqslant}

\title{K-out-of-N Oblivious Transfer over \\Erasure Channels: Capacity and Reductions} \author{\IEEEauthorblockN{Amirhossein Shekofteh and R\'emi~A.~Chou}
\thanks{This work was supported in part by NSF grant CCF-2401373. A preliminary version of this work was presented at the 2026 IEEE International Symposium on Information Theory (ISIT)~\cite{11653775}.}
\thanks{The authors are with the Department of Computer Science and Engineering, The University of Texas at Arlington, Arlington, TX 76019 USA (e-mail: amirhossein.shekofteh@uta.edu; remi.chou@uta.edu).}}
\IEEEoverridecommandlockouts

\maketitle

\begin{abstract}
We study $K$-out-of-$N$ oblivious transfer (OT) between a server and a client over a discrete memoryless channel, under an honest-but-curious model. The client must retrieve $K$ out of $N$ messages stored by the server while learning nothing about the remaining $N-K$ messages, and the server must not learn the client's selection. For a binary erasure channel (BEC) with parameter $\epsilon$, we show that the $K$-out-of-$N$ OT capacity is $\min\Big(\frac{K\epsilon}{N-K},1-\epsilon\Big)$. This characterization generalizes the known capacity results for $1$-out-of-$2$ OT and for $1$-out-of-$N$ OT when $\epsilon\ge\tfrac{N-1}{N}$. We also investigate OT reductions, specifically, we show that constructing a $K$-out-of-$N$ OT protocol from repeating $K$ times an optimal $1$-out-of-$N$ OT protocol  can be suboptimal, but we also identify certain ranges of $\epsilon$ and $M < N$ for which a $1$-out-of-$N$ OT protocol can be optimally realized via the repetition of a $1$-out-of-$M$ OT protocol.
 Finally, we discuss extensions of our results to a generalized erasure channel (GEC).
\end{abstract}
\section{Introduction}
Oblivious transfer (OT)~\cite{rabin2005exchange,even1985randomized}  is a fundamental primitive for secure two-party computation~\cite{kilian1988founding}.  In $K$-out-of-$N$ OT, the server stores $N$ messages and the client must exactly recover $K$ of them so that $(i)$ the client learns no information about the remaining $N-K$ messages and $(ii)$ the server learns no information about which $K$ messages were requested. \par

It is well known that information-theoretically secure OT is impossible with a single server and noiseless communication alone. One approach to overcoming this limitation relies on data replication across multiple servers and shared randomness among them, e.g.,~\cite{9599633,sun2018capacity}. A separate line of work considers  
  settings where a noisy channel between participants is the resource enabling information-theoretic security~\cite{crepeau1988achieving,crepeau1997efficient,crepeau2004efficient,4036292,nascimento2008oblivious,ahlswede2007oblivious,Ahlswede2013,10619607,11195360,shekofteh2025improved,11122510,5513725,5205944,7377066,10737288,7817799,10735295c,11195437,10735295}. For 1-out-of-2 OT,~\cite{ahlswede2007oblivious} derives capacity bounds for the generalized erasure channel (GEC), and establishes the capacity for the binary erasure channel (BEC).~\cite{10619607,11195360} further improve lower bounds over the binary symmetric channel and binary erasure symmetric channel.
For $1$-out-of-$N$ OT,~\cite{4036292} introduces the OT capacity of noisy resources,~\cite{nascimento2008oblivious} derives this capacity for a certain class of channels, and~\cite{shekofteh2025improved} gives an achievable scheme over a BEC with erasure probability $\epsilon$ that improves previously known achievable rates for $\epsilon\in[1/2,1]$.
Previous work also considers other OT settings over noisy channels, including malicious participants~\cite{nascimento2008oblivious,dowsley2017oblivious,pinto2011achieving}, Gaussian and wireless channels~\cite{11122510,5513725,7377066,5205944}, wiretap channels~\cite{10737288,7817799}, and multi-participant OT over multiple-access or broadcast channels~\cite{11195437,10735295,chou2024dual,chou2025dual,chou2021pairwise,aghaee2025networkoblivioustransfernoisy}. To the best of our knowledge, $K$-out-of-$N$ OT from a single server over noisy channels has remained an open problem in the literature. While a straightforward approach to this setting is to repeat a $1$-out-of-$N$ OT protocol $K$ times,  our results show that this strategy is suboptimal.
\par
In this paper, we study $K$-out-of-$N$ OT over a BEC and a GEC, both with parameter $\epsilon$, and a public noiseless channel under honest-but-curious participants. The server stores $N$ files, controls the input of the erasure channel, and the client observes its output. The client and server may exchange messages over the public noiseless channel. The client chooses $K$ files to receive, and the protocol guarantees that the server learns no information about which files were requested and that the client learns no information about the remaining $N-K$ files.
Our main contributions are: $(i)$ the characterization of the $K$-out-of-$N$ OT
capacity over the BEC; $(ii)$ an optimal reduction of a $1$-out-of-$N$ OT protocol to $\frac{N-1}{M-1}$ repeated uses of a $1$-out-of-$M$ OT protocol whenever $M \in \{2, \dots, N-1\}$ satisfies $M \ge \frac{1}{1-\epsilon}$ and $M-1$ divides $N-1$;  and $(iii)$ an achievable rate over the GEC that yields the OT capacity whenever $\epsilon\ge \frac{N-K}{N}$.  \par
Compared to previous work for 1-out-of-2 and $1$-out-of-$N$ OT~\cite{ahlswede2007oblivious,nascimento2008oblivious,shekofteh2025improved,chou2024dual,chou2025dual,chou2021pairwise}, our $K$-out-of-$N$ OT setting introduces new challenges in determining optimal reductions and in establishing the achievability and converse bounds: 
\begin{itemize}
    \item Reduction: A first challenge is determining whether reducing the problem to an  oblivious transfer protocol with a smaller database size, e.g., a 1-out-of-2 OT, is optimal or suboptimal. Previous work~\cite{shekofteh2025improved,chou2025dual} has shown that such reductions are optimal in some cases, in particular when $\epsilon\le1/2$ and $K=1$. 
    \item Achievability bounds: In the $1$-out-of-$N$ OT schemes~\cite{ahlswede2007oblivious,nascimento2008oblivious,shekofteh2025improved}, the server first sends symbols over the noisy channel. Based on the noisy-channel outputs observed by the client, the client forms $N$ index sets and sends them to the server.  The server then uses the transmitted symbols indexed by these sets to mask the $N$ files with a one-time pad.   In the $1$-out-of-$N$ case, only one of these index sets enables recovery of the requested file. In our $K$-out-of-$N$ setting, $K$ index sets must enable recovery of the $K$ requested files, which creates two challenges: how these index sets are created and how they are assigned to the requested files.
    \begin{itemize}
        \item Index-set creation: The $K$ index sets that enable recovery of the requested files are generated jointly, which can create dependencies absent in the $1$-out-of-$N$ OT schemes~\cite{ahlswede2007oblivious,nascimento2008oblivious,shekofteh2025improved}. We design a coding scheme that makes these jointly generated index sets indistinguishable from the remaining index sets.
    \item Index-set assignment: In the $1$-out-of-$N$ OT schemes~\cite{ahlswede2007oblivious,nascimento2008oblivious,shekofteh2025improved}, only one index set enables recovery, so its assignment to the requested file is unique. In our $K$-out-of-$N$ scheme, the $K$ index sets that enable recovery can be assigned to the $K$ requested files in any order. Unless these sets are statistically interchangeable, the assignment can reveal the requested files to the server. We construct index sets whose joint distribution is invariant under any permutation of their labels. 
\end{itemize}
    \item Converse bounds: Previous work~\cite{ahlswede2007oblivious,nascimento2008oblivious,shekofteh2025improved,chou2024dual,chou2025dual,chou2021pairwise}   addresses the case $K=1$, where the converse shows that public communication alone reveals negligible information about a requested file, because that same file must remain hidden when a different file is requested and client privacy makes these requests indistinguishable to the server. For general $K$, the challenge is to show that public communication reveals negligible information about all requested files jointly. Extending the preceding argument would require an alternative request excluding all these files, which is impossible when $K>N-K$. Applying the argument separately to each file is also insufficient, since this can miss relationships revealed by public communication.  Our converse resolves this challenge by showing that public communication reveals negligible information about the requested files jointly, including their relationships.
\end{itemize}

We next compare our results more specifically with those of~\cite{ahlswede2007oblivious,nascimento2008oblivious,shekofteh2025improved}. For 1-out-of-2 OT,~\cite{ahlswede2007oblivious} characterizes the capacity over a BEC with parameter $\epsilon$ as $\min(\epsilon,1-\epsilon)$, and our scheme recovers this result for $N=2$ and $K=1$.   
For $1$-out-of-$N$ OT over a BEC with erasure probability $\epsilon$,~\cite[Proposition~14]{nascimento2008oblivious} shows that the capacity is $1-\epsilon$ when $\frac{N-1}{N}< \epsilon$. Their argument is driven by the requirement that there are enough erased observations to hide the other $N-1$ files, which yields the capacity only in that high-erasure regime. In contrast, our scheme recovers this result and establishes the capacity for any value of $\epsilon$ and any parameters $(N,K)$. In~\cite{shekofteh2025improved}, the scheme first constructs a 1-out-of-$M$ OT protocol, where $M\le N$ is determined by the BEC parameter $\epsilon$ as $M  \triangleq \begin{cases}
        2 & \text{if }  \epsilon\le\frac{1}{2}, \\
        \left\lfloor \frac{1}{1-\epsilon} \right\rfloor  & \text{if }  \frac{1}{2}<\epsilon <1
    \end{cases}$, and subsequently combines several such instances to achieve $1$-out-of-$N$ OT. Our approach instead designs the scheme directly for $1$-out-of-$N$ OT, and more generally for $K$-out-of-$N$ OT, using both erased and non-erased channel observations without such a reduction. As a result, for $K=1$ and arbitrary $N$, our scheme achieves higher rates than~\cite{shekofteh2025improved} over a wide range of $\epsilon$, showing that this reduction approach is suboptimal in general. Finally, we note that a naive $K$-out-of-$N$ scheme can be obtained by repeating the optimal $1$-out-of-$N$ scheme $K$ times, however, our results show that this strategy is suboptimal. \par

We formally describe the problem statement in Section~\ref{sec:PrS}. We state our main results in Section~\ref{sec:MR}.  We present our achievability scheme in Section~\ref{sec:achievability_thm1} and establish our converse in Section~\ref{sec:Converse}. 
Finally, we provide concluding remarks in Section~\ref{sec:concluding}.
\section{Problem Statement} \label{sec:PrS}
For $a, b \in \mathbb{R}$, define
$\llbracket a : b \rrbracket \triangleq [\lfloor a \rfloor , \lceil b \rceil] \cap \mathbb{N}$, and $[a] \triangleq \llbracket1 : a\rrbracket$. We formalize a $K$-out-of-$N$ OT protocol as follows.

\begin{definition} \label{def1}
An $(n, k,K,N)$ oblivious transfer (OT) protocol over a binary erasure channel (BEC) consists of 
\begin{itemize}
        \item $N$ independent strings, $F_{[N]}\triangleq (F_{i})_{i \in [N]}$, where each string is uniformly distributed over $\{0,1\}^k$ and stored on the server;
        \item An integer $K \in [N-1]$, denoting the number of files requested by the client;
        \item A random demand set $\mathcal{Z}$ uniformly chosen over $\mathfrak{Z}\triangleq\{\mathcal{S}\subseteq [N]:|\mathcal{S}|=K\}$, representing the client's choices;
	\item Two random variables $R_0,R_1$, which represent local randomness available at the client and the server, respectively;
	\item For $t \in [n]$, $r_{t}$ denotes the number of public communication rounds between the server and the client between the $t$-th and $(t + 1)$-th uses of the channel,
\end{itemize}
where $F_{[N]}, \mathcal{Z}, R_0, R_1$ are mutually independent, and operates as follows from time $t=1$ to $t=n$:

At time $t$, the server sends $ X_t\in \mathcal{X} \triangleq \{ 0,1 \}$ over the BEC  and the client observes $ Y_t \in \mathcal{Y} \triangleq \{0,1,e\} $, where $X_t (F_{[N]},R_1,(U_{i})_{i \in [t-1]})$\label{1} and $ U_{t}\triangleq(M_{1,t,j},M_{2,t,j})_{j \in [r_{t}]}$ denotes all the messages publicly exchanged between the server and the client between the $t$-th and $(t + 1)$-th uses of the channel. For each time $t$, the messages are generated iteratively from $j=1$ to $j=r_t$ as follows:
			\begin{itemize}
			\item The server sends to the client the message $M_{1,t,j}$, which is a function of
				\begin{align}
					(&F_{[N]}, R_1, (U_{s})_{s\in [t-1]}, (M_{2,t,s})_{s\in[j-1]}), \label{Eq: M1} 
				\end{align}
			\item The client sends to the server the message $M_{2,t,j}$, which is a function of 
            \begin{align}
		\!\!(\mathcal{Z}, R_0, Y^t\!, (U_{s})_{s\in [t-1]}, (M_{1,t,i})_{i \in [j]}),\label{Eq: M2}\end{align}
				\end{itemize}
where $Y^t\triangleq(Y_1,\ldots,Y_t)$. Define the entire public communication as $U\triangleq (U_{t})_{t \in [n]}$. Finally, the client forms $ \widehat{F}_{\mathcal{Z}} $, an estimate of $F_{\mathcal{Z}}$, from $(\mathcal{Z} , R_0 , Y^n, U)$. 	
\end{definition}
\begin{definition}\label{D2}
The rate $R$ is achievable if there exists a sequence of $(n,k,K,N)$ OT protocols  such that 
\begin{align}
 \lim_{n\to\infty} \frac{Kk}{n} &=R,\nonumber\\
\lim_{n\to\infty} \mathbb{P}\big[\widehat{F}_{\mathcal{Z}} \neq F_{\mathcal{Z}}\big] &= 0, \label{Eq: C1} \\
\lim_{n\to\infty} I\big(\mathcal{Z}; F_{[N]}, R_1, X^n, U \big) &= 0, \label{Eq: C2} \\
\lim_{n\to\infty} I\big(F_{[N]\setminus \mathcal{Z}}; U, Y^n , \mathcal{Z}, R_0\big) &= 0. \label{Eq: C3}
\end{align}
Define the supremum of all achievable rates as the OT capacity~$C^{\binom{N}{K}}$. 
\end{definition}
Equation~\eqref{Eq: C1} ensures that the client obtains the selected files. 
Equation~\eqref{Eq: C2} ensures the privacy of the client by keeping the selection of files $\mathcal{Z}$ private from the server. Equation~\eqref{Eq: C3} ensures the privacy of the server by keeping non-selected files private from the client.

\section{Main Results}\label{sec:MR}
  Section~\ref{MR-subsec1} states the capacity of $K$-out-of-$N$ OT over a BEC. Section~\ref{MR-subsec2} studies a reduction-based achievability scheme and identifies the parameter regimes in which it is optimal. Finally, Section~\ref{MR-subsec3} extends the achievability analysis to a generalized erasure channel (GEC) with  parameter $\epsilon$ and derives the OT capacity when $\epsilon \ge \frac{N-K}{N}$.
\subsection{Capacity result for the BEC}\label{MR-subsec1}
Theorem~\ref{Theorem_1_Capacity} characterizes the capacity of $K$-out-of-$N$ oblivious transfer over the BEC and recovers known capacity results as special cases. We use examples to highlight key properties of the capacity expression and compare it with the naive approach that repeats an optimal $1$-out-of-$N$ protocol $K$ times.
\begin{theorem}\label{Theorem_1_Capacity}
Let $N\ge2$, $K\in[N-1]$, and consider a BEC with parameter $\epsilon \in [0,1]$. The $K$-out-of-$N$ OT capacity~is 
\begin{align}
    C^{\binom{N}{K}}=\min\Bigg(\frac{K\epsilon}{N-K},1-\epsilon\Bigg).\nonumber
\end{align}
    \end{theorem}
\begin{proof}
        See Sections~\ref{sec:achievability_thm1} and~\ref{sec:Converse} for the achievability and converse, respectively.  
    \end{proof}
\begin{corollary}\label{cor:naiv_Nacimento}
Let  $N \ge 2$, $K=1$, and consider a BEC with parameter $\epsilon \in [0,1]$. Then, Theorem~\ref{Theorem_1_Capacity} gives the $1$-out-of-$N$ OT capacity
\begin{align}
  C^{\binom{N}{1}} = \min\Bigg(\frac{\epsilon}{N-1},\,1-\epsilon\Bigg).\nonumber
\end{align}
Furthermore, for $\epsilon \ge \frac{N-1}{N}$, we have $C^{\binom{N}{1}} = 1-\epsilon$, which recovers~\cite[Proposition~14]{nascimento2008oblivious}.
\end{corollary}
\begin{corollary}\label{cor:ahlswede}
Let $N=2$, $K=1$, and consider a BEC with parameter $\epsilon\in[0,1]$.
Then,  Theorem~\ref{Theorem_1_Capacity} gives the   $1$-out-of-$2$ OT capacity $C^{\binom{2}{1}}=\min(\epsilon,\,1-\epsilon)$, which recovers~\cite[Theorem~1]{ahlswede2007oblivious}.
\end{corollary}
\begin{remark}\label{EX:compare_with_naive}
   The naive approach that repeats an optimal $1$-out-of-$N$ OT protocol $K$ times to build a $K$-out-of-$N$ OT protocol yields the rate $R_N=C^{\binom{N}{1}}$. This rate is suboptimal in general, and its gap to $C^{\binom{N}{K}}$  is 
\begin{align}
C^{\binom{N}{K}}-R_N=
\begin{cases}
\dfrac{N(K-1)\epsilon}{(N-1)(N-K)},  &0\le \epsilon\le \dfrac{N-K}{N}, \nonumber\\[1.2ex]
1-\dfrac{N\epsilon}{N-1}, & \dfrac{N-K}{N}<\epsilon<\dfrac{N-1}{N}, \nonumber\\[1.2ex]
0, & \epsilon\ge \dfrac{N-1}{N}.
\end{cases}
\end{align}

\end{remark}

\begin{figure}[!ht]
\centering
\subfloat[OT capacity $C^{\binom{30}{K}}$ versus $K$ for
  $\epsilon=0.4$. The capacity increases with $K$ up to $K^\star=18$ and
  then saturates at $1-\epsilon = 0.6$.\label{fig:capacity_vs_K}]
  {\begin{minipage}[t]{0.41\linewidth}\vspace{0pt}\centering
    \includegraphics[width=\linewidth]{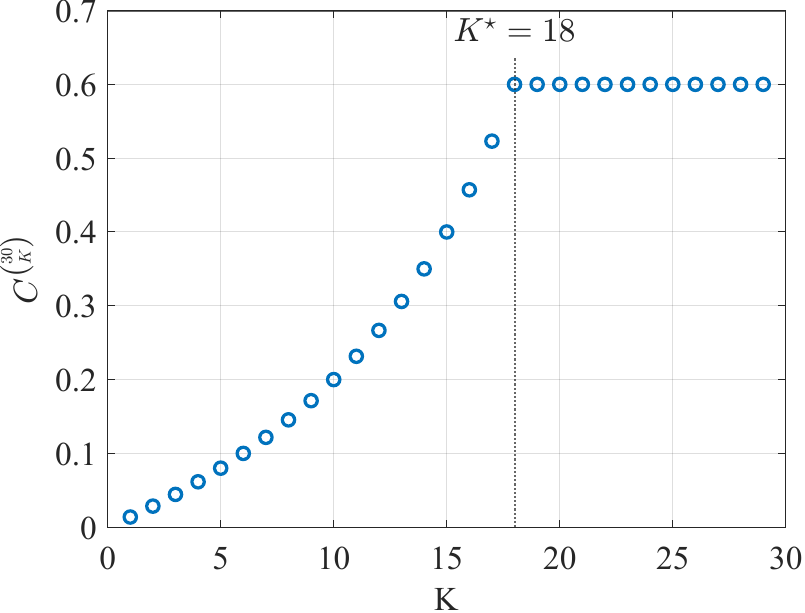}
  \end{minipage}}
\qquad \qquad
\subfloat[OT capacity  $C^{\binom{10}{6}}$ versus $\epsilon$. The capacity increases linearly up to
  $\epsilon^\star = 0.4$ and then decreases linearly, with maximum
  $\frac{K}{N}=0.6$. \label{fig:capacity_vs_epsilon}]
    {\begin{minipage}[t]{0.41\linewidth}\vspace{-3pt}\centering
    \includegraphics[width=\linewidth]{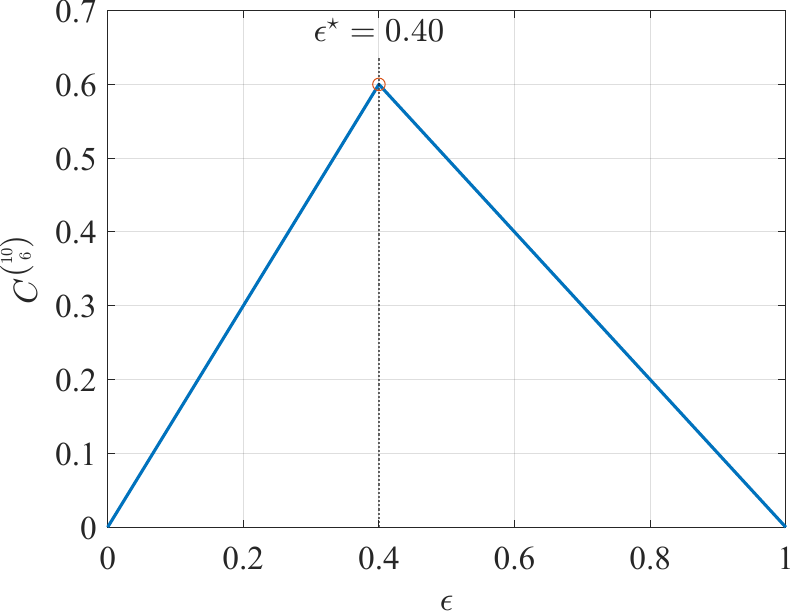}
  \end{minipage}}
\caption{ Subfigure~(a) corresponds to Example~\ref{ex:graph_fixed_epsilon};
subfigure~(b) corresponds to Example~\ref{ex:graph_fixed_K}.}
\end{figure}

\begin{properties}
Consider a BEC with parameter $\epsilon \in [0,1]$.
From Theorem~\ref{Theorem_1_Capacity}, the $K$-out-of-$N$ OT capacity has the following properties:
\begin{enumerate}
    \item \label{Pr:ahlswede}
Let $N = \alpha K$, where $\alpha \in \mathbb{N}$, and $\alpha \geq 2$. Then, the OT capacity is 
\begin{align}
  C^{\binom{\alpha K}{K}}=\min\Bigg(\frac{\epsilon}{\alpha-1},1-\epsilon\Bigg).\nonumber
\end{align}
For $\alpha = 2$,   $C^{\binom{2K}{K}} = \min(\epsilon,1-\epsilon)$. Corollary~\ref{cor:ahlswede} is the special case $K=1$, $N=2$.
    \item \label{Pr:Kstar} For fixed $\epsilon$ with $\frac{1}{N}\le \epsilon<1$, define $K^\star\triangleq \left\lceil N(1-\epsilon)\right\rceil$. The OT capacity in Theorem~\ref{Theorem_1_Capacity} increases with $K$  up to $K^\star$ and remains constant at $1-\epsilon$ for $K^\star \le K < N$.
    \item \label{Pr:Nacimento} For $\epsilon \geq \frac{N-K}{N}$, the OT capacity in Theorem~\ref{Theorem_1_Capacity} reduces to $C^{\binom{N}{K}} = 1 - \epsilon$.~\cite[Proposition~14]{nascimento2008oblivious}  corresponds to the special case $K=1$.
    \item \label{Pr:estar}
For fixed $N$ and $K$, let $\epsilon^\star = \frac{N-K}{N}$.  
Then, $C^{\binom{N}{K}}(\epsilon)$ increases linearly on $[0,\epsilon^\star]$ and decreases
linearly on $[\epsilon^\star,1]$. The maximum value is reached at
$\epsilon = \epsilon^\star$ and equals
$C^{\binom{N}{K}}(\epsilon^\star) = \frac{K}{N}$.
\end{enumerate}
\end{properties}
\begin{example}\label{ex:graph_fixed_epsilon}
This example illustrates Property~\ref{Pr:Kstar} for fixed $N$ and $\epsilon$. Let $N = 30$ and $\epsilon = 0.4$. Then, from Property~\ref{Pr:Kstar}, as shown in Fig.~\ref{fig:capacity_vs_K}, the capacity increases with $K$  up to $K^\star = N(1-\epsilon) = 18$. After this point, the capacity stays constant at $0.6=1-\epsilon$ as $K$ increases.
\end{example}

\begin{example}\label{ex:graph_fixed_K}
This example illustrates Property~\ref{Pr:estar} for fixed $N$ and $K$. Let $N = 10$ and $K = 6$. Then, from
Property~\ref{Pr:estar}, as shown in Fig.~\ref{fig:capacity_vs_epsilon},
the capacity increases linearly with $\epsilon$ up to $\epsilon^\star = 0.4$, where the maximum is $\frac{K}{N} =\frac{6}{10}$. After this point, the capacity decreases linearly to zero at $\epsilon=1$.
\end{example}

\begin{figure}[!ht]
\centering
{
  \begin{minipage}[t]{0.47\linewidth}\vspace{0pt}\centering    \includegraphics[width=\linewidth]{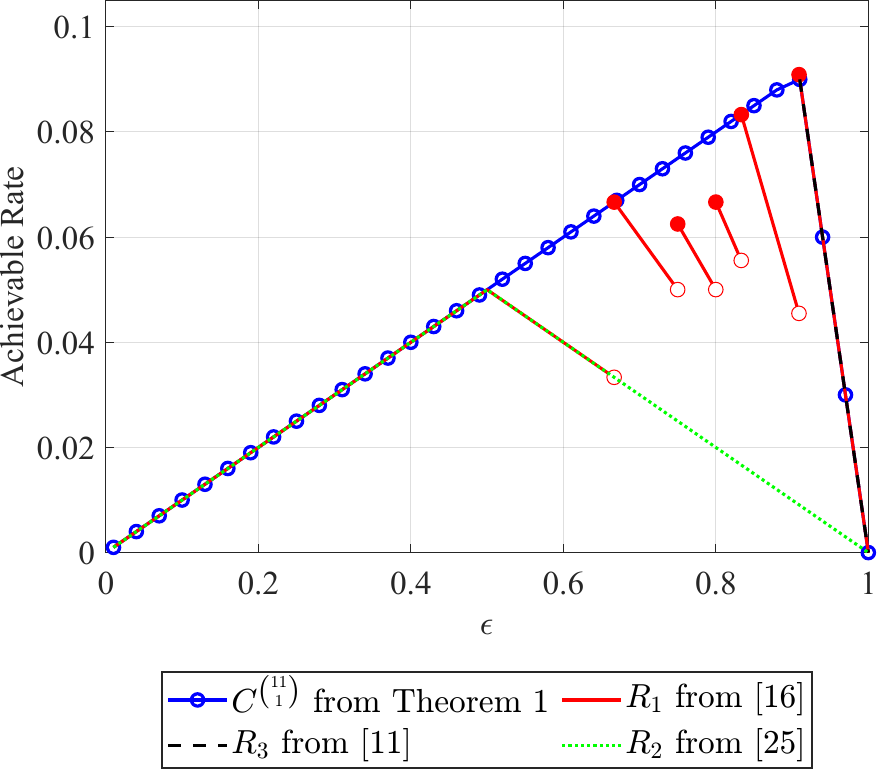}
  \end{minipage}
}

\caption{For a BEC with parameter $\epsilon$ and $(K,N)=(1,11)$, we compare the capacity in Theorem~\ref{Theorem_1_Capacity}, denoted by $C^{\binom{11}{1}}$, with the achievable rate in~\cite{shekofteh2025improved}, denoted by $R_1$, the achievable rate of the single-server configuration in~\cite{10735295}, denoted by $R_2$, and the achievable rate in~\cite[Prop.~14]{nascimento2008oblivious}, denoted by $R_3$. At the discontinuity points of $R_2$, filled red circles indicate included endpoint values, while empty red circles indicate excluded endpoint values.\label{fig:rate_comparison}}
\end{figure}

 \begin{example}\label{ex:rate_comparison2}
In this example, we compare our result for the special case $K=1$ with previously known results. Specifically, for $(K,N)=(1,11)$, Fig.~\ref{fig:rate_comparison} compares the capacity in Theorem~\ref{Theorem_1_Capacity}, denoted by $C^{\binom{11}{1}}$, with $(i)$ the achievable rate from~\cite{shekofteh2025improved}, denoted by $R_{1}$, $(ii)$ the achievable rate from~\cite[Theorem~1]{10735295} for the special case of a single server, denoted by $R_2$, and $(iii)$ the achievable rate from~\cite[Proposition~14]{nascimento2008oblivious}, denoted by $R_3$. Note that $R_3$ is only defined when $\epsilon\geq \frac{N-1}{N}$.
\end{example}

\subsection{Reduction-based achievable rate}\label{MR-subsec2}

Protocol reductions   construct a $1$-out-of-$N$ OT protocol  by repeated use of a  $1$-out-of-$M$ OT protocol~\cite{10.1007/3-540-48910-X_4}, where $M<N$. In practice, this means that a system that already supports $1$-out-of-$M$ OT can handle a larger library of $N$ files without changing its underlying architecture.
Unlike~\cite{10.1007/3-540-48910-X_4}, where the $1$-out-of-$M$ OT protocol is treated as a black box with no associated rate, we assume here that it is constructed from a BEC and has a rate as defined in Definition~\ref{D2}.  Consequently, for each choice of $M$, the reduction induces a corresponding $1$-out-of-$N$ OT protocol with an achievable rate, denoted by $R_{\mathrm{red}}$.
 Our objective is then to determine, as a function of $\epsilon$,  the values of $M$   for which the reduction is optimal, in the sense that the induced rate $R_{\mathrm{red}}$ equals the $1$-out-of-$N$ OT capacity $C^{\binom{N}{1}}$.

\begin{theorem}\label{Theorem_2_def}
 Let $N\ge 3$ and consider a BEC with parameter $\epsilon\in[0,1]$. For any integer $M$ satisfying $2\le M<N$, suppose a $1$-out-of-$M$ OT protocol with rate $C^{\binom{M}{1}}$  is given.  Then, there exists a reduction that makes  $\left\lceil \frac{N-1}{M-1}\right\rceil$ repeated uses of the $1$-out-of-$M$ OT protocol and yields a $1$-out-of-$N$ OT protocol with achievable rate
\begin{align}
R_{\mathrm{red}}(M)=\frac{C^{\binom{M}{1}}}{\left\lceil \frac{N-1}{M-1}\right\rceil }.\nonumber
\end{align}
Moreover, the reduction is optimal, i.e., $R_{\mathrm{red}}(M)=C^{\binom{N}{1}},$ for every $M\in\mathcal{P}_\epsilon(N)$, where $$\mathcal{P}_\epsilon(N) \triangleq \left\{M\in\{2,\dots,N-1\}: \epsilon\le \frac{M-1}{M}, \ \text{and } M-1 | N-1 \right\}.$$ 
\end{theorem}
\begin{proof}
    See Appendix~\ref{sec:Achievability of Theorem 1}.
\end{proof}

\begin{corollary}
By Theorem~\ref{Theorem_2_def}, for $\epsilon \in [0,\frac{1}{2}]$,  $M=2$ yields an optimal reduction for every $N\ge 3$.
\end{corollary}

\begin{figure}[!ht]
\centering {
  \begin{minipage}[t]{0.49\linewidth}\vspace{0pt}\centering
    \includegraphics[width=\linewidth]{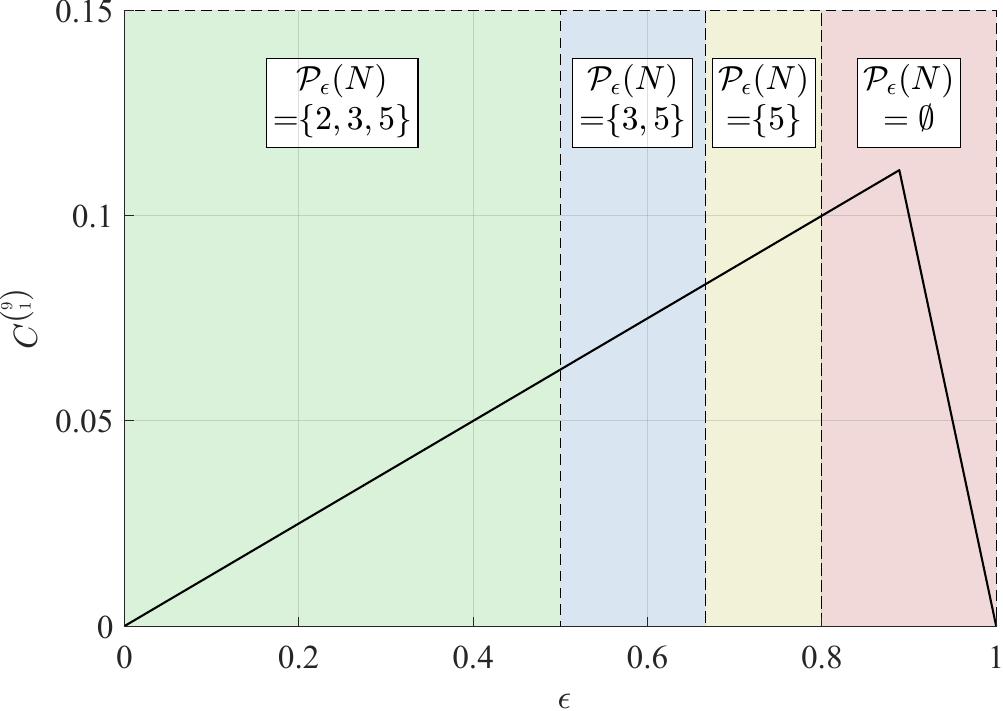}
  \end{minipage}
}
\caption{For  $(K,N)=(1,9)$, the black curve shows the capacity $C^{\binom{9}{1}}$ from Theorem~\ref{Theorem_1_Capacity} as a function of the BEC parameter $\epsilon$. The shaded regions indicate the corresponding sets $\mathcal{P}_\epsilon(9)$. As $\epsilon$ increases, the set $\mathcal{P}_\epsilon(9)$ shrinks from $\{2,3,5\}$ to $\emptyset$.}\label{fig:validM}
\end{figure}

\begin{example}\label{Ex:MergedM}
Consider a BEC with parameter $\epsilon\in[0,1]$. For a fixed $N$, let $\mathcal P_0(N)\triangleq \{M\in\{2,\dots,N-1\}: M-1| N-1\}$. The set $\mathcal P_\epsilon(N)$ is nonincreasing in $\epsilon$ and is obtained by removing from $\mathcal P_0(N)$ each $M \in \mathcal P_0(N)$ once $\epsilon>\frac{M-1}{M}$. 
Figure~\ref{fig:validM} illustrates this for $N=9$, and Table~\ref{Table:ValidM} lists the evolution of $\mathcal P_{\epsilon}(N)$ with $\epsilon$ for various $N$.
For $N=9$, since $N-1=8$, we have $\mathcal P_0(9)=\{2,3,5\}$, and, as $\epsilon$ increases,   $\mathcal P_{\epsilon}(9)$ loses $\{2\}$, $\{3\}$, and $\{5\}$  at $\epsilon=\frac12$, $\epsilon=\frac23$, and $\epsilon=\frac45$,~respectively.
\end{example}

\begin{table*}[t]
\centering
\caption{Set $\mathcal{P}_0(N)$ and evolution of $\mathcal{P}_\epsilon(N)$ as $\epsilon$ increases. 
}
\label{Table:ValidM}
\renewcommand{\arraystretch}{1.3}
\resizebox{\textwidth}{!}{
\begin{tabular}{c|c|c|c|c|c|c|c|c}
\hline
\multirow{2}{*}{$N$} & \multirow{2}{*}{$\mathcal{P}_0(N)$} & \multicolumn{7}{c}{$\mathcal{P}_\epsilon(N)$} \\ \cline{3-9} 
 & & $0 \le \epsilon \le \frac{1}{2}$ & $\frac{1}{2} < \epsilon \le \frac{2}{3}$ & $\frac{2}{3} < \epsilon \le \frac{3}{4}$ & $\frac{3}{4} < \epsilon \le \frac{4}{5}$ & $\frac{4}{5} < \epsilon \le \frac{5}{6}$ & $\frac{5}{6} < \epsilon \le \frac{6}{7}$ & $\frac{6}{7} < \epsilon \le 1$ \\ \hline
5  & $\{2,3\}$         & $\{2,3\}$         & $\{3\}$         & $\emptyset$ & $\emptyset$ & $\emptyset$ & $\emptyset$ & $\emptyset$ \\
9  & $\{2,3,5\}$       & $\{2,3,5\}$       & $\{3,5\}$       & $\{5\}$       & $\{5\}$       & $\emptyset$ & $\emptyset$ & $\emptyset$ \\
13 & $\{2,3,4,5,7\}$   & $\{2,3,4,5,7\}$   & $\{3,4,5,7\}$   & $\{4,5,7\}$   & $\{5,7\}$   & $\{7\}$       & $\{7\}$       & $\emptyset$ \\
16 & $\{2,4,6\}$       & $\{2,4,6\}$       & $\{4,6\}$       & $\{4,6\}$       & $\{6\}$       & $\{6\}$       & $\emptyset$ & $\emptyset$ \\
20 & $\{2\}$           & $\{2\}$           & $\emptyset$ & $\emptyset$ & $\emptyset$ & $\emptyset$ & $\emptyset$ & $\emptyset$ \\ \hline
\end{tabular}
}
\end{table*}

\subsection{Achievable rate for the GEC}\label{MR-subsec3}

\begin{definition}\label{DEf3}
\cite[Definition~2]{Ahlswede2013} A GEC is a discrete memoryless channel  $( \mathcal{X},W_{Y|X },\mathcal{Y})$, where, for some nonempty set $\mathcal{Y}_1 \subset \mathcal{Y}$, the probabilities $W_{Y| X} (y|x)$, $y \in \mathcal{Y}_1$, do not depend on $x \in \mathcal{X}$. Outputs $y \in \mathcal{Y}_1$ are considered erasures as they provide no information about the inputs. The parameter  of a GEC is defined as $\epsilon \triangleq \sum_{y \in \mathcal{Y}_1} W(y| x)$.
\end{definition}
We extend our results to a GEC with parameter $\epsilon$. First, we derive an achievable rate for $K$-out-of-$N$ OT over a GEC and identify a parameter regime for which we characterize   the $K$-out-of-$N$ OT capacity over the GEC, denoted by $C_{\mathrm{GEC}}^{\binom{N}{K}}$.

\begin{theorem}\label{Theorem:GEC_achievable_Rate}
   Consider a GEC $(\mathcal X,W_{Y|X},\mathcal Y)$ with parameter $0<\epsilon<1$. 
   For $N\ge 2$ and $K\in[N-1]$, an  achievable rate is
  \begin{align}
      R_{\mathrm{GEC}}=\frac{\max_{P_X}I(X;Y)}{1-\epsilon}\min\Big(\frac{K\epsilon}{N-K},1-\epsilon\Big).\label{achievablerate_GEC}
  \end{align} 
\end{theorem}
\begin{proof}
    See Appendix~\ref{Sec:ach_for_GEC}.
\end{proof}

\begin{corollary}\label{Cor:GEC}
Under the notation of Theorem~\ref{Theorem:GEC_achievable_Rate}, 
 we have 
\begin{align}
     \begin{cases}
        C_{\mathrm{GEC}}^{\binom{N}{K}}\ge \frac{K\epsilon}{(N-K)(1-\epsilon)}\max_{P_X}I(X;Y), & \epsilon< \tfrac{N-K}{N},\nonumber\\[4pt]
        C_{\mathrm{GEC}}^{\binom{N}{K}}=\max_{P_X}I(X;Y), & \epsilon\ge \tfrac{N-K}{N},
    \end{cases}
  \end{align}
which, for $K=1$ and  $\epsilon\ge \tfrac{N-K}{N}$, recovers~\cite[Proposition~14]{nascimento2008oblivious}.
\end{corollary}
\begin{proof}
    See Appendix~\ref{Sec:ProofcorGEC}.
\end{proof}

\begin{corollary}\label{cor:recover_AhlswedeCsiszar_T3}
Under the notation of Theorem~\ref{Theorem:GEC_achievable_Rate}, for $N=2$ and $K=1$, we have
\begin{align}
\begin{cases}
C_{\mathrm{GEC}}^{\binom{2}{1}}\ge \frac{\epsilon}{1-\epsilon}\max_{P_X}I(X;Y), & \epsilon< \tfrac{1}{2},\nonumber\\[4pt]
C_{\mathrm{GEC}}^{\binom{2}{1}}=\max_{P_X}I(X;Y), & \epsilon\ge \tfrac{1}{2},
\end{cases}  
\end{align}
which recovers~\cite[Theorem~3]{ahlswede2007oblivious}.
\end{corollary}
\section{Achievability of Theorem~\ref{Theorem_1_Capacity}} \label{sec:achievability_thm1}

Section~\ref{subsec:coding_ach1} introduces the coding scheme, Section~\ref{Sec:clientPrivacyAGK-BEC} proves client privacy, Section~\ref{Subsec:serverPrivacy_th1} proves server privacy, and  Section~\ref{subs:ach_rate_th1} derives the achievable rate.

\subsection{Coding scheme}\label{subsec:coding_ach1}
The coding scheme is described in Algorithm~\ref{AGK-BEC}. The main idea of Algorithm~\ref{AGK-BEC} uses the pure passing and pure erasing behavior of the BEC, i.e., when $Y_t\in\{0,1\}$ the client learns $X_t$, while when $Y_t=e$ it learns nothing about the channel input $X_t$. The client constructs $N$ pairwise disjoint index sets that are identical in size and distribution from the server's point of view. To ensure the client obtains exactly the $K$ files while preserving the client privacy, $K$ sets will come from non-erased positions. To preserve the server privacy, $N-K$ sets will come from erased positions. 
The non-erased part contains about $(1-\epsilon)n$ positions to form $K$ sets, and the erased part contains about $n\epsilon$ positions to form $N-K$ sets, which motivates the introduction of the following quantity to define the size of the $N$ sets:
\begin{align} 
r \triangleq \min\left(\frac{1-\epsilon}{K},\frac{\epsilon}{N-K}\right). \label{eq:rdef}
\end{align} 
Then, the client labels positions in  Line~\ref{Line: AGK-BEC Line 4}  to ensure that non-erased positions are labeled  over $[K]$ and erased positions are labeled over $[\![K+1:N]\!]$. This labeling yields $N$ preliminary sets. The client then in Line~\ref{Line AGK-BEC Nline 3} trims each preliminary set to the same size, applies a permutation $\phi$ to ensure $K$ sets that come from non-erased positions are assigned to the client's requested files, i.e., $\phi(\mathcal Z)=[K]$, and sends the resulting $N$ sets to the server. 
Finally, the server applies its channel inputs over the $N$ sets as one-time pads to its files and sends the results to the client. For the $K$ sets drawn from non-erased positions, the client can cancel the pad and recover the requested files. For the $N-K$ sets drawn from erased positions, the pad remains unknown to the client.

Algorithm~\ref{AGK-BEC} is illustrated in Example~\ref{Ex:AGK-BEC_example} below.

\begin{algorithm}    
\caption{$K$-out-of-$N$ OT Scheme} \label{AGK-BEC} 
\begin{algorithmic}[1]
 \setcounter{ALG@line}{0}
\State The server sends a random binary sequence $X^n \triangleq (X_t)_{t \in [n]}$ over a BEC with parameter $\epsilon$. \label{Line: AGK-BEC Line 1}

\State Upon observing $Y^n$, the client proceeds as follows:\label{Line: AGK-BEC Line 2}
\begin{algsubstates}
  \State Define \label{Line: AGK-BEC Line 3} 
      \begin{align}
      \mathcal{G} \triangleq \{ t \in [n] : Y_t \in \{0,1\} \},\quad 
      \mathcal{B} \triangleq \{ t \in [n] : Y_t = e \}.\nonumber
      \end{align}
    \State For each $t\in[n]$, generate a label $J_t\in[\![0:N]\!]$ as follows: \label{Line: AGK-BEC Line 4}
    \begin{align}
    \!\!\!\!\!\!\!\!\!\mathbb{P}[J_t=j | t\in\mathcal G] &=
    \begin{cases}\nonumber
    1-\frac{Kr}{1-\epsilon}, & j=0,\\[2pt]
    \frac{r}{1-\epsilon}, & j\in[K],\\[2pt]
    0, & j\in\![\![K+1:N]\!],
    \end{cases}\\[6pt]
    \!\!\!\!\!\!\!\!\!\mathbb{P}[J_t=j | t\in\mathcal B] &=
    \begin{cases}\nonumber
    1-\frac{(N-K)r}{\epsilon}, & j=0,\\[2pt]
    \frac{r}{\epsilon}, & j\in\![\![K+1:N]\!],\\[2pt]
    0, & j\in[K],
    \end{cases}
    \end{align}
    where $r$ is defined in~\eqref{eq:rdef}. $J_t=0$ means that index $t$ is discarded.
\State Define \label{Line: AGK-BEC Line 5}
\begin{align}
        \tilde{\mathcal C}_u \triangleq \{t\in[n]: J_t=u\},\qquad u\in[N].\nonumber
\end{align}
\end{algsubstates}
  \State \label{Line AGK-BEC Nline 3}
  Define 
  \begin{align}
    m\triangleq\Big\lfloor(r-\frac{\delta}{K})n\Big\rfloor\label{eq:mdef},
    \end{align}
where $\delta>0$ and $r$ is defined in~\eqref{eq:rdef}. If
$|\tilde{\mathcal C}_u|<m$ for some $u\in[N]$, then the client aborts the
protocol. Otherwise, the client proceeds as follows:
  \begin{algsubstates}
  \State   For each $u\in[N]$, independently sample $\mathcal C_u$ uniformly at random from all subsets of $\tilde{\mathcal C}_u$ of size $m$. \label{Line: AGK-BEC Line 6}
  \State For $s\in[N]$, define\label{Line: AGK-BEC Line 7}
      \begin{align}
        \mathcal L_s \triangleq \mathcal C_{\phi(s)},\nonumber
      \end{align}
      where $\phi:[N]\to[N]$ is any bijection such that $\phi(\mathcal Z)=[K]$.
      \State Send $\mathcal{L}_{[N]}\triangleq (\mathcal{L}_i)_{i\in [N]}$ to the server over the public channel.\label{Line: AGK-BEC Line 8}
    \end{algsubstates}
\State The server sends $(X^n[\mathcal L_s]\oplus F_s)_{s\in[N]} $ over the public channel. \label{Line: AGK-BEC Line 9}
\State The client recovers $F_s$ by computing $(X^n[\mathcal L_s]\oplus F_s)\oplus X^n[\mathcal L_s]=F_s$, for each $s\in\mathcal{Z}$.\label{Line: AGK-BEC Line 10}
\end{algorithmic}
\end{algorithm}

\begin{example}\label{Ex:AGK-BEC_example}
Let $\epsilon=\frac{3}{4}$, $N=4$, $K=2$, and $\mathcal Z=\{2,3\}$. Then, $r=\min\Big(\frac{1-\epsilon}{K},\frac{\epsilon}{N-K}\Big)=\frac{1}{8}$.  The server sends $X^n$ and the client receives $Y^n$. The client then forms the sets $\mathcal G$ and $\mathcal B$ as in Line~\ref{Line: AGK-BEC Line 3}. For each $t\in\mathcal G$, the client generates a label $J_t$ according to Line~\ref{Line: AGK-BEC Line 4},
$ \mathbb P[J_t=u| t\in\mathcal G]=\frac{r}{1-\epsilon}=\frac{1}{2}, u\in\{1,2\}, $ 
and $ \mathbb P[J_t=0| t\in\mathcal G]=0. $ 
For each $t\in\mathcal B$, the client generates a label $J_t$ according to
$\mathbb P[J_t=u| t\in\mathcal B]=\frac{r}{\epsilon}=\frac{1}{6}, u\in\{3,4\},$ 
and
$\mathbb P[J_t=0| t\in\mathcal B]=1-\frac{(N-K)r}{\epsilon}=\frac{2}{3}.$ 
This produces the  sets $\tilde{\mathcal C}_1,\tilde{\mathcal C}_2,\tilde{\mathcal C}_3,\tilde{\mathcal C}_4$ in Line~\ref{Line: AGK-BEC Line 5}.
If each preliminary set has size at least
$ 
m=\left\lfloor \left(r-\frac{\delta}{K}\right)n\right\rfloor,
$ 
the client samples subsets $\mathcal C_u\subseteq \tilde{\mathcal C}_u$ of size $m$ as in Line~\ref{Line: AGK-BEC Line 6}.
Next, the client chooses a bijection $\phi:[4]\to[4]$ such that $\phi(\mathcal Z)=[2]$. For example, the client may choose 
$\phi(2)=1$, $\phi(3)=2$, $\phi(1)=3$, $\phi(4)=4.$ 
Then, by Line~\ref{Line: AGK-BEC Line 7},
$\mathcal L_1=\mathcal C_3$, $\mathcal L_2=\mathcal C_1$, $\mathcal L_3=\mathcal C_2$, $ \mathcal L_4=\mathcal C_4.$ 
The client sends $(\mathcal L_s)_{s\in[4]}$ to the server, and the server returns $(X^n[\mathcal L_s]\oplus F_s)_{s\in[4]}$.
Because $\mathcal L_2$ and $\mathcal L_3$ are formed from non-erased positions, the client knows $X^n[\mathcal L_2]$ and $X^n[\mathcal L_3]$ and recovers $F_2$ and $F_3$. In contrast, $\mathcal L_1$ and $\mathcal L_4$ are formed from erased positions, so the client obtains no information about $F_1$ and $F_4$.
\end{example}

\subsection{Client privacy: Ensuring privacy of file selection}\label{Sec:clientPrivacyAGK-BEC}
We  show that the server gains no information about the client's choice by proving~\eqref{Eq: C2} and considering that the only local randomness used by the server is $X^n$.  We first present a series of lemmas.
\begin{lemma}\label{Lemma:Prob_r}
For any fixed $t \in [n]$ and any $u \in [N]$, we have
\begin{align}
  \mathbb{P}[t \in \tilde{\mathcal C}_u] = r,  \nonumber
\end{align}
where $\tilde{\mathcal{C}}_u$  is defined in Line~\ref{Line: AGK-BEC Line 5} of Algorithm~\ref{AGK-BEC}.
\end{lemma}
\begin{proof} For $\tilde{\mathcal{C}}_u$ defined in Line~\ref{Line: AGK-BEC Line 5} of Algorithm~\ref{AGK-BEC}, we  have 
    \begin{align} 
    \mathbb{P}[t \in \tilde{\mathcal{C}}_u]
    &\stackrel{(a)}{=}\mathbb{P}[t \in \tilde{\mathcal{C}}_u,\, t\in\mathcal{G}]+ \mathbb{P}[t \in \tilde{\mathcal{C}}_u,\, t\in\mathcal{B}]\nonumber\\
    & \stackrel{(b)}{=} \mathbb{P}[t \in \mathcal{G}]\mathbb{P}[J_t=u | t \in \mathcal{G}] + \mathbb{P}[t \in \mathcal{B}]\mathbb{P}[J_t=u | t \in \mathcal{B}] \nonumber\\
    &\stackrel{(c)}{=} (1-\epsilon)\mathbb{P}[J_t=u | t \in \mathcal{G}] + \epsilon\mathbb{P}[J_t=u | t \in \mathcal{B}] \nonumber\\
    &\stackrel{(d)}{=} 
\begin{cases} 
(1-\epsilon)\frac{r}{1-\epsilon}+\epsilon \cdot 0  & \text{if } u \in [K], \nonumber\\
 (1-\epsilon)\cdot 0+\epsilon \frac{r}{\epsilon} & \text{if } u \in [\![K+1, N]\!],
\end{cases} \nonumber\\
&= r,\nonumber
\end{align}
where $(a)$ holds because  $\mathcal G$ and $\mathcal B$ partition $[n]$,   $(b)$ holds because for $t\in\mathcal G$ and $t\in\mathcal B$ the event $\{t\in\tilde{\mathcal C}_u\}$ occurs if and only if $J_t=u$, $(c)$ holds by Line~\ref{Line: AGK-BEC Line 3} of Algorithm~\ref{AGK-BEC}, and $(d)$ holds by Line~\ref{Line: AGK-BEC Line 4} of Algorithm~\ref{AGK-BEC}.
\end{proof}

\begin{lemma}\label{Lemma:ExchangeablePreTrimmed}
For any permutation $\phi$ of $[N]$ and any pairwise disjoint subsets $\ell_1,\ldots,\ell_N\subseteq[n]$, 
\begin{align}
\mathbb{P}[\tilde{\mathcal C}_{\phi(1)}=\ell_1,\ldots,\tilde{\mathcal C}_{\phi(N)}=\ell_N]
=\mathbb{P}[\tilde{\mathcal C}_{1}=\ell_1,\ldots,\tilde{\mathcal C}_{N}=\ell_N].\nonumber\end{align}
\end{lemma}
\begin{proof}
Define $\ell_0 \triangleq [n]\setminus \bigcup_{i=1}^N \ell_i$. Then, we have
\begin{align}
\mathbb{P}[\tilde{\mathcal C}_{\phi(1)}=\ell_1,\ldots,\tilde{\mathcal C}_{\phi(N)}=\ell_N] 
&=
\mathbb{P}\!\Big[\bigcap_{i=1}^N\bigcap_{t\in \ell_i}\{t\in\tilde{\mathcal C}_{\phi(i)}\}\ \cap\ \bigcap_{t\in \ell_0}\{t\notin\tilde{\mathcal C}_{[N]}\}\Big] \nonumber\\
&\stackrel{(a)}{=}
\prod_{i=1}^N\prod_{t\in \ell_i}\mathbb{P}[t\in\tilde{\mathcal C}_{\phi(i)}]\ \cdot\ \prod_{t\in \ell_0}\mathbb{P}[t\notin\tilde{\mathcal C}_{[N]}] \nonumber\\
&\stackrel{(b)}{=}
\prod_{i=1}^N\prod_{t\in \ell_i}\mathbb{P}[t\in\tilde{\mathcal C}_{i}]\ \cdot\ \prod_{t\in \ell_0}\mathbb{P}[t\notin\tilde{\mathcal C}_{[N]}] \nonumber\\
&=
\mathbb{P}[\tilde{\mathcal C}_{1}=\ell_1,\ldots,\tilde{\mathcal C}_{N}=\ell_N],\nonumber\end{align}
where $(a)$ holds because the BEC is memoryless and the label $J_t$  is generated independently for each $t$ in Line~\ref{Line: AGK-BEC Line 4} of Algorithm~\ref{AGK-BEC}, and $(b)$ holds by Lemma~\ref{Lemma:Prob_r}.
\end{proof}
\begin{lemma}\label{Lemma:PermInvTrimmed}
For any permutation $\phi$ of $[N]$ and any pairwise disjoint subsets
$\ell_1,\ldots,\ell_N\subseteq[n]$,
\begin{align}
\mathbb{P}[\mathcal{C}_{\phi(1)}\!=\!\ell_1,\ldots,\mathcal{C}_{\phi(N)}=\ell_N ]
=\mathbb{P}[\mathcal{C}_{1}=\ell_1,\ldots,\mathcal{C}_{N}=\ell_N ].\nonumber\end{align}
\end{lemma}
\begin{proof}
Define $\mathfrak T_m\triangleq\Big\{(\tilde{\ell}_u)_{u\in[N]}:\tilde{\ell}_u\subseteq[n],|\tilde{\ell}_u|\ge m~\forall u\in[N],
\tilde{\ell}_1,\ldots,\tilde{\ell}_N\text{ are pairwise disjoint}\Big\}.$ Then, for any $\ell_1,\ldots,\ell_N\subseteq[n]$, we have
\begin{align}
\mathbb{P}[\mathcal{C}_{\phi(1)}=\ell_1,\ldots,\mathcal{C}_{\phi(N)}=\ell_N ] \nonumber
&\stackrel{(a)}{=}
\!\!\!\!\!\!\!\!\!\!\sum_{\substack{(\tilde{\ell}_u)_{u\in[N]}\in\mathfrak T_m}}\!\!\!\!\!\!\!\!\!\!\!\Big(
\mathbb{P}[(\tilde{\mathcal{C}}_{\phi(u)}\!\!=\!\tilde{\ell}_u)_{ u\in[N]}]\mathbb{P}[(\mathcal{C}_{\phi(u)}\!\!=\!\ell_u)_{u\in[N]} | (\tilde{\mathcal{C}}_{\phi(u)}\!\!=\!\tilde{\ell}_u)_{u\in[N]}] \Big)\nonumber\\
&\stackrel{(b)}{=}\!\!\!\!\!\!\!\!\!\!
\sum_{\substack{(\tilde{\ell}_u)_{u\in[N]}\in\mathfrak T_m}}\!\!\!\!\!\!\!\!\!\!
\mathbb{P}[(\tilde{\mathcal{C}}_{\phi(u)}=\tilde{\ell}_u)_{u\in[N]}]\,
\prod_{i=1}^N \frac{\mathds{1}\{\ell_i\subseteq \tilde{\ell}_i,\ |\ell_i|=m\}}{\binom{|\tilde{\ell}_i|}{m}} \nonumber\\
&\stackrel{(c)}{=}\!\!\!\!\!\!\!\!\!\!
\sum_{\substack{(\tilde{\ell}_u)_{u\in[N]}\in\mathfrak T_m}}\!\!\!\!\!\!\!\!\!\!
\mathbb{P}[(\tilde{\mathcal{C}}_{u}=\tilde{\ell}_u)_{u\in[N]}]\,
\prod_{i=1}^N \frac{\mathds{1}\{\ell_i\subseteq \tilde{\ell}_i,\ |\ell_i|=m\}}{\binom{|\tilde{\ell}_i|}{m}} \nonumber\\
&=
\mathbb{P}[\mathcal{C}_{1}=\ell_1,\ldots,\mathcal{C}_{N}=\ell_N ],\nonumber
\end{align}
where $(a)$ holds by the law of total probability, $(b)$ holds by Line~\ref{Line: AGK-BEC Line 6} of Algorithm~\ref{AGK-BEC}, and $(c)$ holds by Lemma~\ref{Lemma:ExchangeablePreTrimmed}. 
\end{proof}
\begin{lemma}\label{Lemma:MIprivacy}
For $\mathcal{L}_{[N]}$ defined in Line~\ref{Line: AGK-BEC Line 8} of Algorithm~\ref{AGK-BEC}, we have
\begin{align}
I(\mathcal L_{[N]};\mathcal Z)=0.\nonumber
\end{align}
\end{lemma}
\begin{proof}
For $\mathcal{S}\in\mathfrak{Z}$, the bijection $\phi_{\mathcal{S}}:[N]\to[N]$ in Line~\ref{Line: AGK-BEC Line 7}, and any $\ell_1,\ldots,\ell_N\subseteq[n]$ with $|\ell_i|=m$, where $m$ is defined in~\eqref{eq:mdef}, we have 
\begin{align}
\mathbb{P}[\mathcal L_{1}=\ell_1,\ldots,\mathcal L_{N}=\ell_N|\mathcal{Z}=\mathcal{S}]
&\stackrel{(a)}{=}\mathbb{P}[\mathcal C_{\phi(1)}=\ell_1,\ldots,\mathcal C_{\phi(N)}=\ell_N | \mathcal{Z}=\mathcal{S}] \nonumber\\
&\stackrel{(b)}{=}
\mathbb{P}[\mathcal C_{\phi(1)}=\ell_1,\ldots,\mathcal C_{\phi(N)}=\ell_N] \nonumber\\
&\stackrel{(c)}{=}
\mathbb{P}[\mathcal C_{1}=\ell_1,\ldots,\mathcal C_{N}=\ell_N],\label{firstpart}
\end{align}
where $(a)$ holds by Line~\ref{Line: AGK-BEC Line 7} of Algorithm~\ref{AGK-BEC}, $(b)$ holds because the construction of $(\mathcal{C}_u)_{u\in[N]}$ depends only on $(Y^n,R_0)$ and is independent of $\mathcal{Z}$, and $(c)$ holds by Lemma~\ref{Lemma:PermInvTrimmed}. Then, we have
\begin{align}
\mathbb P[\mathcal L_1=\ell_1,\ldots,\mathcal L_N=\ell_N]  \nonumber
&=
\sum_{\mathcal{S}\in\mathfrak Z}\mathbb P[\mathcal{Z}=\mathcal{S}]\,
\mathbb P[\mathcal L_1=\ell_1,\ldots,\mathcal L_N=\ell_N | \mathcal{Z}=\mathcal{S}] \nonumber\\
&=\sum_{\mathcal{S}\in\mathfrak Z}\mathbb P[\mathcal Z=\mathcal{S}]\,
\mathbb P[\mathcal C_1=\ell_1,\ldots,\mathcal C_N=\ell_N] \nonumber\\
&=
\mathbb P[\mathcal C_1=\ell_1,\ldots,\mathcal C_N=\ell_N], \label{eq:marg-law-final}
\end{align}
where the second equality holds by~\eqref{firstpart}. Then, from \eqref{firstpart} and \eqref{eq:marg-law-final}, for every $\mathcal{S}\in\mathfrak Z$, we have
$\mathbb P[\mathcal L_1=\ell_1,\ldots,\mathcal L_N=\ell_N | \mathcal Z=\mathcal{S}]
=\mathbb P[\mathcal L_1=\ell_1,\ldots,\mathcal L_N=\ell_N]$.
\end{proof}
We now show that the full view of the server reveals no information about the client's choice set. Specifically, we have
\begin{align} 
    I(F_{[N]},X^n,U;\mathcal{Z})
    &\stackrel{(a)}{=}I(F_{[N]},X^n,(F_{i}\oplus X^n[\mathcal{L}_i])_{i\in [N]},\mathcal{L}_{[N]};\mathcal{Z})\nonumber\\
    &\stackrel{(b)}{=}I(F_{[N]},X^n,\mathcal{L}_{[N]};\mathcal{Z})\nonumber\\
    &=I(X^n,\mathcal{L}_{[N]};\mathcal{Z})+I(F_{[N]};\mathcal{Z}|X^n,\mathcal{L}_{[N]})\nonumber\\
    &\le I(X^n,\mathcal{L}_{[N]};\mathcal{Z})+I(F_{[N]};\mathcal{Z},X^n,\mathcal{L}_{[N]})\nonumber\\
    &\stackrel{(c)}{=}I(X^n,\mathcal{L}_{[N]};\mathcal{Z})\nonumber\\
    &=I(\mathcal{L}_{[N]};\mathcal{Z})+I(X^n;\mathcal{Z}|\mathcal{L}_{[N]})\nonumber\\
    &\stackrel{(d)}{=}I(X^n;\mathcal{Z}|\mathcal{L}_{[N]})\nonumber\\
    &\leq I(X^n;\mathcal{Z},R_0,\mathcal{L}_{[N]},\mathcal{G},\mathcal{B})\nonumber\\
    &\stackrel{(e)}{=}I(X^n;\mathcal{Z},R_0,\mathcal{G},\mathcal{B})\nonumber\\
    &\stackrel{(f)}{=}I(X^n;\mathcal{G},\mathcal{B})\nonumber\\
    &\stackrel{(g)}{=}0, \label{eq:clientprivacyAGK-BEC}
\end{align}
where $(a)$ holds by Lines~\ref{Line: AGK-BEC Line 8}--\ref{Line: AGK-BEC Line 9} of Algorithm~\ref{AGK-BEC}, $(b)$ holds because $(F_{i} \oplus X^n[\mathcal{L}_i])_{i \in [N]}$ is a function of $(F_{[N]}, X^n, \mathcal{L}_{[N]})$,
 $(c)$  holds because $F_{[N]}$ is independent of $\mathcal{Z}$, $X^{n}$, and $\mathcal{L}_{[N]}$, $(d)$ follows from Lemma~\ref{Lemma:MIprivacy}, $(e)$ holds because $\mathcal{L}_{[N]}$ is a function of  $(\mathcal{Z},R_0,\mathcal{G},\mathcal{B})$ by Lines~\ref{Line: AGK-BEC Line 2}--\ref{Line AGK-BEC Nline 3} of Algorithm~\ref{AGK-BEC}, $(f)$ holds because $\mathcal{Z}$ and $R_0$  are independent of $(X^n,\mathcal{G},\mathcal{B})$, and $(g)$ holds because $X^n$ is independent of  $\mathcal{G}$ and $\mathcal{B}$.     
\subsection{Server privacy: Ensuring privacy of unselected files}\label{Subsec:serverPrivacy_th1}
We show that the client gains no information about unselected files by proving~\eqref{Eq: C3}.
We first show that, for any unselected file $F_s$, the corresponding pad $X^n[\mathcal L_s]$ remains independent of the client's side information, even when all other files are revealed.

\begin{lemma}\label{PCA32}
For any $\mathcal S\in\mathfrak Z$ and $s\in[N]\setminus\mathcal S$, let $\mathcal L_i$, $i\in[N]$, be defined in Line~\ref{Line: AGK-BEC Line 7} of Algorithm~\ref{AGK-BEC}, and define $\mathcal L'\triangleq[N]\setminus\{s\}$. Then, we have
\begin{align}
I\big(
X^n[\mathcal L_s];
Y^n,(F_i\oplus X^n[\mathcal L_i])_{i\in\mathcal L'},R_0,F_{[N]}
| \mathcal Z=\mathcal S
\big)=0.
\nonumber
\end{align}
\end{lemma}
\begin{proof}
Fix $\mathcal{S} \in \mathfrak{Z}$ and $s \in [N]\setminus\mathcal{S}$. Then, we have
\begin{align}
    &I(X^n[\mathcal{L}_s];Y^n,(F_{i}\oplus X^n[\mathcal{L}_i])_{i\in\mathcal{L}'},F_{[N]},R_0|\mathcal{Z}=\mathcal{S})\nonumber\\
    &=I(X^n[\mathcal{L}_s];Y^n,( X^n[\mathcal{L}_i])_{i\in\mathcal{L}'},F_{[N]},R_0|\mathcal{Z}=\mathcal{S})\nonumber\\
    &=I(X^n[\mathcal{L}_s];Y^n|\mathcal{Z}=\mathcal{S})\nonumber +I(X^n[\mathcal{L}_s];( X^n[\mathcal{L}_i])_{i\in\mathcal{L}'},F_{[N]},R_0|Y^n,\mathcal{Z}=\mathcal{S})\nonumber\\
    &\stackrel{(a)}{=}I(X^n[\mathcal{L}_s];( X^n[\mathcal{L}_i])_{i\in\mathcal{L}'},F_{[N]},R_0|Y^n,\mathcal{Z}=\mathcal{S})\nonumber\\
    &=I(X^n[\mathcal{L}_s];R_0|Y^n,\mathcal{Z}=\mathcal{S})+I(X^n[\mathcal{L}_s];( X^n[\mathcal{L}_i])_{i\in\mathcal{L}'},F_{[N]}|R_0,Y^n,\mathcal{Z}=\mathcal{S})\nonumber\\    
    &\stackrel{(b)}=I(X^n[\mathcal{L}_s];( X^n[\mathcal{L}_i])_{i\in\mathcal{L}'}|R_0,Y^n,\mathcal{Z}=\mathcal{S})+I(X^n[\mathcal{L}_s];F_{[N]}|( X^n[\mathcal{L}_i])_{i\in\mathcal{L}'},R_0,Y^n,\mathcal{Z}=\mathcal{S})\nonumber\\
    &\stackrel{(c)}{=} I(X^n[\mathcal{L}_s];( X^n[\mathcal{L}_i])_{i\in\mathcal{L}'}|R_0,Y^n,\mathcal{Z}=\mathcal{S})\nonumber\\
    &\stackrel{(d)}{=}0, \nonumber
    \end{align}
where $(a)$ holds because, for $s\in[N]\setminus\mathcal S$, we have
$\mathcal L_s\subseteq\mathcal B$, and therefore
$X^n[\mathcal L_s]$ is independent of $Y^n$ conditioned on
$\mathcal Z=\mathcal S$, $(b)$ holds by the chain rule and because erased positions at $X^n[\mathcal L_s]$ are i.i.d. and independent of $R_0$  conditioned on $(Y^n,\mathcal Z=\mathcal S)$, $(c)$ holds because
$F_{[N]}$ is independent of $(X^n,Y^n,R_0,\mathcal Z)$, and
$(d)$ holds because, conditioned on $(Y^n,R_0,\mathcal Z=\mathcal S)$, the sets $\mathcal L_{[N]}$ are fixed and pairwise disjoint, so the i.i.d. generation of $X^n$ and the memorylessness of the BEC imply that $X^n[\mathcal L_s]$ and $(X^n[\mathcal L_i])_{i\in\mathcal L'}$ are independent.
\end{proof}

We next recall a one-time-pad lemma that will be used to establish server privacy.
\begin{lemma}[{\cite{chou2024dual}}] \label{Lemma 10}
    Let $\lambda, \delta > 0$. Consider the random variables $A$, $B$, $C$, $D$ such that
$C$ and $D$ are defined over $\{0, 1\}^m$, $H(D) > m - \delta$, and $I(D; A,B,C) \leq \lambda$. Then, we have
\begin{align}
I(A; B, C \oplus D) \leq I(A; B) + \lambda + \delta. \nonumber
\end{align}
\end{lemma}

We now use Lemma~\ref{PCA32} and Lemma~\ref{Lemma 10} to conclude the proof of server privacy. We have
\begin{align}
I(F_{[N]\setminus\mathcal{Z}};U,Y^n,\mathcal{Z},R_0)
&\stackrel{(a)}{=}I(F_{[N]\setminus\mathcal{Z}};(F_{i}\oplus X^n[\mathcal{L}_i])_{i\in[N]},\mathcal{L}_{[N]},Y^n,\mathcal{Z},R_0)\nonumber\\
&\stackrel{(b)}{=}I(F_{[N]\setminus\mathcal{Z}};(F_{i}\oplus X^n[\mathcal{L}_i])_{i\in[N]},Y^n,\mathcal{Z},R_0)\nonumber\\
&=I(F_{[N]\setminus\mathcal{Z}};\mathcal{Z})
+I(F_{[N]\setminus\mathcal{Z}};(F_{i}\oplus X^n[\mathcal{L}_i])_{i\in[N]},Y^n,R_0|\mathcal{Z})\nonumber\\
&\stackrel{(c)}{=}I(F_{[N]\setminus\mathcal{Z}};(F_{i}\oplus X^n[\mathcal{L}_i])_{i\in[N]},Y^n,R_0|\mathcal{Z})\nonumber\\
&\stackrel{(d)}{=}\sum_{\mathcal{S}\in\mathfrak{Z}}\mathbb{P}[\mathcal{Z}=\mathcal{S}]
I(F_{[N]\setminus\mathcal{S}};(F_{i}\oplus X^n[\mathcal{L}_i])_{i\in[N]},Y^n,R_0|\mathcal{Z}=\mathcal{S})\nonumber\\
&\stackrel{(e)}{=}\sum_{\mathcal{S}\in\mathfrak{Z}}\mathbb{P}[\mathcal{Z}=\mathcal{S}]\!\!\!
\sum_{j\in[N]\setminus\mathcal{S}}\!\!\!
I(F_j;(F_{i}\oplus X^n[\mathcal{L}_i])_{i\in[N]},Y^n,R_0
| F_{[j-1]\setminus\mathcal{S}},\mathcal{Z}\!=\!\mathcal{S})\nonumber\\
&\leq
\sum_{\mathcal{S}\in\mathfrak{Z}}\mathbb{P}[\mathcal{Z}=\mathcal{S}]\!\!\!
\sum_{j\in[N]\setminus\mathcal{S}}\!\!\!
I(F_j;(F_{i}\oplus X^n[\mathcal{L}_i])_{i\in[N]},Y^n,R_0,
F_{[N]\setminus\{j\}}|\mathcal{Z}=\mathcal{S})\nonumber\\
&\stackrel{(f)}{\leq} \sum_{\mathcal{S}\in\mathfrak{Z}}\mathbb{P}[\mathcal{Z}=\mathcal{S}]\!\!\!
\sum_{j\in[N]\setminus\mathcal{S}}\!\!\!
\!\![I(F_j;(F_{i}\oplus X^n[\mathcal{L}_i])_{i\in[N]\setminus\{j\}},Y^n\!\!,R_0,
F_{[N]\setminus\{j\}}|\mathcal{Z}\!=\!\mathcal{S})\!+\!\lambda_{n,j}+\!\delta_n ]\nonumber\\
&\stackrel{(g)}{=}\sum_{\mathcal{S}\in\mathfrak{Z}}\mathbb{P}[\mathcal{Z}=\mathcal{S}]\!\!\!
\sum_{j\in[N]\setminus\mathcal{S}}(\lambda_{n,j}+\delta_n)\nonumber\\
&\xrightarrow{n \to \infty}0,\nonumber
\end{align}
where $(a)$ holds by Lines~\ref{Line: AGK-BEC Line 8}--\ref{Line: AGK-BEC Line 9} of Algorithm~\ref{AGK-BEC}, $(b)$ holds because $\mathcal{L}_{[N]}$ is a function of $Y^n,\mathcal{Z}$, and $R_0$, $(c)$ holds because $\mathcal{Z}$ is independent of the files, $(d)$ holds by conditioning on $\mathcal{Z}$, $(e)$ holds by the chain rule, $(f)$ holds by  applying Lemma~\ref{Lemma 10} under the conditional distribution given $\mathcal Z=\mathcal S$, with $A\leftarrow F_j$, $B\leftarrow B_j$, $C\leftarrow F_j$, and $D\leftarrow X^n[\mathcal L_j]$, using the definitions $B_j \triangleq ((F_i\oplus X^n[\mathcal L_i])_{i\in[N]\setminus\{j\}},Y^n,R_0,F_{[N]\setminus\{j\}})$, $\lambda_{n,j}\triangleq I(X^n[\mathcal L_j];B_j,F_j|\mathcal Z=\mathcal S)$, and $\delta_{n} \triangleq \frac{1}{n}$, $(g)$ holds because files are uniformly distributed and independent of $Y^n$, $R_0$, and $\mathcal{Z}$,  and the limit holds because $\lambda_{n,j}$ vanishes as $n \to \infty$ according to Lemma~\ref{PCA32}, and $\delta_n$ vanishes as $n \to \infty$.
\subsection{Achievable rate}\label{subs:ach_rate_th1}
To study the probability that the algorithm aborts in Line~\ref{Line AGK-BEC Nline 3} of Algorithm~\ref{AGK-BEC}, we first show the following lemma.
\begin{lemma}\label{Lemma:mzero}
For any  $\delta > 0$ and $r$ defined in~\eqref{eq:rdef},  we have 
\begin{align}
    \lim_{n \to \infty} \sum_{u \in [N]}\mathbb{P} \Big[|\tilde{\mathcal C}_u|< n\Big(r - \frac{\delta}{K}\Big)\Big]=0,\nonumber
\end{align}
where $\tilde{\mathcal{C}}_u$ is defined in Line~\ref{Line: AGK-BEC Line 5} of Algorithm~\ref{AGK-BEC}.
\end{lemma}
\begin{proof} 
For  $u \in [N]$, $i \in [n]$, define $X_i^{(u)} \triangleq \mathds{1}\{i \in \tilde{\mathcal{C}}_u\}$.
Due to the memoryless BEC and independent label generation, for each $u \in [N]$, $(X_i^{(u)})_{i\in[n]}$ are independent Bernoulli random variables.
Then, we have 
\begin{align}
\exp\left(-\frac{2n\delta^2}{K^2}\right)
&\stackrel{(a)}\ge \mathbb{P}\Big[\sum_{i=1}^n X_i^{(u)}-\mathbb E\big[\sum_{i=1}^n X_i^{(u)}\big]\le -\frac{n\delta}{K}\Big]\nonumber\\
&\stackrel{(b)}{=}\mathbb{P}\Big[|\tilde{\mathcal C}_u|-nr\le -\frac{n\delta}{K}\Big],\label{eq:hoeff}
\end{align}
where $(a)$ holds by Hoeffding's inequality~\cite{hoeffding1963probability}, and $(b)$ holds because  $|\tilde{\mathcal C}_u|=\sum_{i=1}^n X_i^{(u)}$ and by Lemma~\ref{Lemma:Prob_r}. Then, we have
    \begin{align}
    \sum_{u \in [N]}\mathbb{P}\bigg[|\tilde{\mathcal{C}}_u| < n(r - \frac{\delta}{K})\bigg] \nonumber
    &\le \sum_{u \in [N]} \exp\left(-\frac{2n\delta^2}{K^2}\right)\nonumber\\
    &= N\exp\left(-\frac{2n\delta^2}{K^2}\right)\nonumber\\
    &\xrightarrow{n \to \infty}0,\nonumber
\end{align}
where the inequality holds by~\eqref{eq:hoeff}.
\end{proof}
Following Line~\ref{Line AGK-BEC Nline 3} of Algorithm~\ref{AGK-BEC}, the abort probability satisfies
\begin{align}
\mathbb P\left[
\bigcup_{u\in[N]}
\Big\{|\tilde{\mathcal C}_u|<m\Big\}
\right]
&\stackrel{(a)}{\leq}
\sum_{u\in[N]}
\mathbb P\Big[|\tilde{\mathcal C}_u|<m\Big]\nonumber\\
&\stackrel{(b)}{\leq}
\sum_{u\in[N]}
\mathbb P\Big[|\tilde{\mathcal C}_u|<n\Big(r-\frac{\delta}{K}\Big)\Big]
\nonumber\\
&\xrightarrow{n\to\infty}0,\nonumber
\end{align}
where $(a)$ holds by the union bound, $(b)$ holds because
$m\leq\big(r-\frac{\delta}{K}\big)n$, and the limit holds by
Lemma~\ref{Lemma:mzero}.
Then, on the non-abort event and for $s\in \mathcal{Z}$, the achievable rate is 
\begin{align}
\frac{
|F_{\mathcal{Z}}
|}{n}&=\frac{K|F_s|}{n}\nonumber\\
&\stackrel{(a)}{=}\frac{K|X^n[\mathcal{L}_s]|}{n}\nonumber\\
&=\frac{K|\mathcal{L}_s|}{n}\nonumber\\
&\stackrel{(b)}{=}\frac{Km}{n}\nonumber\\
&\xrightarrow{n \to \infty}K\Big(r-\frac{\delta}{K}\Big)\nonumber\\
&\stackrel{(c)}{=}\min\Big(\frac{K\epsilon}{N-K},1-\epsilon \Big)-\delta,\label{eq:AGK-BEC_ach_rate}
\end{align}
where $(a)$ holds because $ X^n [\mathcal{L}_{s}]$ and $F_ {s} $ have the same size by Line~\ref{Line: AGK-BEC Line 9} of Algorithm~\ref{AGK-BEC}, $(b)$ holds by Line~\ref{Line: AGK-BEC Line 6} of Algorithm~\ref{AGK-BEC}, the  limit holds by~\eqref{eq:mdef}, and $(c)$  holds by~\eqref{eq:rdef}.

\section{Converse of Theorem~\ref{Theorem_1_Capacity}} \label{sec:Converse}

In Section~\ref{sec:MI-bound}, we derive an upper bound on $C^{\binom{N}{K}}$ based on the mutual information between the channel input and output. In Section~\ref{sec:CE-bound}, we derive a second upper bound on $C^{\binom{N}{K}}$ based on the conditional entropy of the channel input given the channel output. By taking the minimum of these two bounds, we obtain
\begin{align}
   C^{\binom{N}{K}}\le\min\Bigg(\max_{p_{X}}I(X;Y),\frac{K}{N-K}\max_{p_{X}}H(X|Y)\Bigg),\label{capacityconversebound}
\end{align}
where the maximization is over the input distribution $p_X$, for the given channel $p_{Y|X}$.
\subsection{Mutual information bound}\label{sec:MI-bound}
To obtain our first  converse bound, we start by deriving the following lemma.  
	\begin{lemma} \label{Lemma:Reduce_U_t}
	For $t \in [n]$, define $U_{[t]}\triangleq (U_i)_{i\in [t]}$. Then, we have  
	\begin{align}
	&I(F_{[N]},R_1; Y^t,R_0 | U_{[t]},\mathcal{Z})\leq I(F_{[N]},R_1; Y^t,R_0 | U_{[t-1]},\mathcal{Z}). \nonumber 
	\end{align}
    \end{lemma}
	\begin{proof} For $t \in [n]$, define $M_{1,t,[r_t]}\triangleq (M_{1,t,j})_{j\in [r_t]}$ and $ M_{2,t,[r_t]}\triangleq (M_{2,t,j})_{j\in [r_t]}$. Then,  we have 
 \begin{align}
		I(F_{[N]},R_1;Y^t,R_0|U_{[t]},\mathcal{Z}) \nonumber  
		&=I(F_{[N]},R_1;Y^t,R_0|U_{[t-1]},M_{1,t,[r_t]},M_{2,t,[r_t]},\mathcal{Z}) \nonumber \\  
		&\stackrel{(a)}{\leq}\!\!I(F_{[N]},R_1\!;\!Y^t\!\!,\!R_0,M_{2,t,r_t}|U_{[t-1]},\!M_{1,t,[r_t]},\!M_{2,t,[r_t-1]},\!\mathcal{Z}) \label{(22)} \\ 
        &= I(F_{[N]},R_1;Y^t,R_0|U_{[t-1]},M_{1,t,[r_t]},M_{2,t,[r_t-1]},\mathcal{Z}) \nonumber \\ 
        & +\!I(F_{[N]},\!R_1\!;\!M_{2,t,r_t}|Y^t\!,R_0,U_{[t-1]},\!M_{1,t,[r_t]},\!M_{2,t,[r_t-1]},\!\mathcal{Z}) \nonumber \\ 
        &\stackrel{(b)}{=} I(F_{[N]},R_1;Y^t,R_0|U_{[t-1]},M_{1,t,[r_t]},M_{2,t,[r_t-1]},\mathcal{Z}) \nonumber \\  
        &\stackrel{(c)}{\leq}\! I(F_{[N]},R_1,M_{1,t,r_t}\!;\!Y^t\!,R_0|U_{[t-1]},\!M_{1,t,[r_t-1]},\!M_{2,t,[r_t-1]},\!\mathcal{Z}) \nonumber \\  
		&= I(F_{[N]},R_1;Y^t,R_0|U_{[t-1]},M_{1,t,[r_t-1]},M_{2,t,[r_t-1]},\mathcal{Z}) \nonumber \\  
        & + \!I(M_{1,t,r_t}; Y^t\!,R_0|F_{[N]},R_1,U_{[t-1]},\!M_{1,t,[r_t-1]},\!M_{2,t,[r_t-1]},\!\mathcal{Z}) \nonumber \\  
		&\stackrel{(d)}{=} I(F_{[N]},R_1;Y^t,R_0|U_{[t-1]},M_{1,t,[r_t-1]},M_{2,t,[r_t-1]},\mathcal{Z}) \label{(23)} \\
		&\stackrel{(e)}{\le} I(F_{[N]},R_1;Y^t,R_0|U_{[t-1]},\mathcal{Z}), \nonumber 
		\end{align}
where $(a)$ holds by the chain rule, $(b)$ holds by~\eqref{Eq: M2}, $(c)$ holds by the chain rule, $(d)$ holds by~\eqref{Eq: M1}, and $(e)$ follows from repeating~\eqref{(22)}-\eqref{(23)} $r_t-1$ times.
\end{proof}
The next lemma applies Lemma~\ref{Lemma:Reduce_U_t} to bound the information  that the channel output $Y^t$ and the client's local randomness  $R_0$ reveal about the requested files $F_{\mathcal Z}$, given  the public communication $U_{[t]}$ up to round $t$ and $\mathcal Z$. 
	\begin{lemma} \label{Lemma:XtYt}
For $t \in [n]$, define $U_{[t]}\triangleq (U_i)_{i\in [t]}$. Then, we have 
\begin{align}
		I(F_{\mathcal{Z}}; Y^t,R_0 | U_{[t]},\mathcal{Z}) \leq \sum_{i=1}^{t}I(X_i;Y_i ). \nonumber
		\end{align}
    \end{lemma}
    \begin{proof}
	For $t \in [n]$, define $U_{[t]}\triangleq (U_i)_{i\in [t]}$. Then, we have 
 \begin{align}
			&I(F_{\mathcal{Z}}; Y^t,R_0 | U_{[t]},\mathcal{Z}) \nonumber\\
            &\leq I(F_{[N]},R_1; Y^t,R_0 | U_{[t]},\mathcal{Z}) \nonumber \\ 
			&\stackrel{(a)}{\leq} I(F_{[N]},R_1; Y^t,R_0| U_{[t-1]},\mathcal{Z}) \label{eq:start} \\  
			&= I(F_{[N]},R_1; Y^{t-1},R_0| U_{[t-1]},\mathcal{Z})+I(F_{[N]},R_1; Y_t| U_{[t-1]},\mathcal{Z},R_0,Y^{t-1}) \nonumber \\  
			 &= I(F_{[N]},\!R_1;\!Y^{t-1}\!\!\!,R_0| U_{[t-1]},\mathcal{Z})\! +\! H(Y_t | U_{[t-1]},\!\mathcal{Z},R_0,Y^{t-1})-\!H(Y_t| U_{[t-1]},\!\mathcal{Z},R_0,Y^{t-1}\!\!\!,F_{[N]},R_1) \nonumber \\  
			&\stackrel{(b)}{\leq} I(F_{[N]},\!R_1; Y^{t-1},R_0| U_{[t-1]},\mathcal{Z})\nonumber  + H(Y_t)-\!H(Y_t| U_{[t-1]},\!\mathcal{Z},R_0,Y^{t-1}\!\!\!,F_{[N]},R_1)\nonumber\\
            &\stackrel{(c)}{\leq} I(F_{[N]},R_1; Y^{t-1},R_0| U_{[t-1]},\mathcal{Z})\nonumber  + H(Y_t) -H(Y_t | X_t, U_{[t-1]},\mathcal{Z},R_0,Y^{t-1},F_{[N]},R_1) \nonumber \\  
			&\stackrel{(d)}{\leq} I(F_{[N]},R_1; Y^{t-1},R_0| U_{[t-1]},\mathcal{Z}) + H(Y_t)-H(Y_t | X_t) \nonumber \\  
			&= I(F_{[N]},R_1; Y^{t-1},R_0| U_{[t-1]},\mathcal{Z})  + I(Y_t;X_t) \label{eq:end} \\  
			&\stackrel{(e)}{\leq} \sum_{i=1}^{t}{I(X_i;Y_i )}, \nonumber
		\end{align}
where $(a)$ follows from Lemma~\ref{Lemma:Reduce_U_t}, $(b)$ holds because conditioning reduces entropy, $(c)$ holds because $X_t$ is a function of $(F_{[N]},R_1,U_{[t-1]})$, $(d)$ holds because $(U_{[t-1]},\mathcal{Z},R_0,Y^{t-1},F_{[N]},R_1)-X_t-Y_t$ forms a Markov chain, and $(e)$ follows from repeating~\eqref{eq:start}-\eqref{eq:end} $t-1$ times. 
\end{proof}

The next lemma shows that if  $W$ satisfies $\lim_{n\to\infty} I(F_z, W; \mathcal{Z}) = 0$ for a fixed file $F_z$, then  the conditional entropy of $F_z$ given $W$ is asymptotically the same for any two realizations of $\mathcal Z$.
\begin{lemma}\label{Lemma:Swap}
For $z\in [N]$, consider a random variable $W$ such that $\lim_{n\to \infty} I(F_z, W; \mathcal{Z}) = 0$. Then, for $\mathcal{S}, \mathcal{S}'\in \mathfrak{Z}$, we have  
\begin{align}
    &H(F_{z} | W  , \mathcal{Z}=\mathcal{S})= H(F_{z} |  W  , \mathcal{Z}=\mathcal{S}') 
    + o(n). \nonumber
\end{align}
\end{lemma}
\begin{proof}
See Appendix~\ref{app:C}.\end{proof}

The next lemma applies Lemma~\ref{Lemma:Swap} to establish that the public communication \(U\) reveals asymptotically no information about the requested files $F_{\mathcal Z}$ given  $\mathcal Z$.
\begin{lemma}\label{Lemma:MISwap}
    We have
    \begin{align}
        I(F_{\mathcal{Z}};U|\mathcal{Z})=o(n).\nonumber
    \end{align}
\end{lemma}
\begin{proof}
For $\mathcal{S}\in \mathfrak{Z}$, write  $\mathcal{S}=\{z_1,\ldots,z_K\}$, and for $i \in [K]$, define $\ell_i'\triangleq(\mathcal{S}\setminus\{z_i\})\cup\{z'\}$ with $z'\in [N]\setminus \mathcal{S}$. Then, we have
    \begin{align}
        I(F_{\mathcal{Z}};U|\mathcal{Z}=\mathcal{S})
        &\stackrel{(a)}{=}\sum_{i=1}^K I(F_{z_i};U|F_{z_{[i-1]}},\mathcal{Z}=\mathcal{S})\nonumber\\
        &=\sum_{i=1}^K H(F_{z_i}|F_{z_{[i-1]}},\mathcal{Z}=\mathcal{S})-H(F_{z_i}|U,F_{z_{[i-1]}},\mathcal{Z}=\mathcal{S})\nonumber\\
        &\stackrel{(b)}{\le}\sum_{i=1}^K H(F_{z_i}|\mathcal{Z}=\ell_i')-H(F_{z_i}|U,F_{z_{[i-1]}},\mathcal{Z}=\mathcal{S})\nonumber\\
        &\stackrel{(c)}{=}\sum_{i=1}^K H(F_{z_i}|\mathcal{Z}=\ell_i')-H(F_{z_i}|U,F_{z_{[i-1]}},\mathcal{Z}=\ell_i')+o(n)\nonumber\\
        &\stackrel{(d)}{\le}\sum_{i=1}^K  H(F_{z_i}|\mathcal{Z}=\ell_i')\nonumber -H(F_{z_i}|Y^n,R_0,U,F_{z_{[i-1]}},\mathcal{Z}=\ell_i')+o(n)\nonumber\\
        &=\sum_{i=1}^K I(F_{z_i};Y^n,R_0,U,F_{z_{[i-1]}}|\mathcal{Z}=\ell_i')+o(n)\nonumber\\
        &=\sum_{i=1}^K \Big(I(F_{z_i};Y^n,R_0,U|\mathcal{Z}=\ell_i') +I(F_{z_i};F_{z_{[i-1]}}|Y^n,R_0,U,\mathcal{Z}=\ell_i')\Big)+o(n)\nonumber\\
        &\stackrel{(e)}{=}\sum_{i=1}^K \Big(I(F_{z_i};Y^n,R_0,U|\mathcal{Z}=\ell_i') +I(F_{z_i};F_{z_{[i-1]}}|\hat{F}_{z_{[i-1]}},Y^n,R_0,U,\mathcal{Z}=\ell_i')\Big)+o(n)\nonumber\\
        &\stackrel{(f)}{\le}\sum_{i=1}^K I(F_{z_i};U,Y^n,R_0|\mathcal{Z}=\ell_i')+o(n)\nonumber\\
        &\stackrel{(g)}{=}o(n),\label{eq:laston}
    \end{align}
    where $(a)$ holds by the chain rule with $F_{z_{[i-1]}}\triangleq(F_{z_{j}})_{j \in [i-1]}$, $(b)$ holds because conditioning reduces entropy and $F_{z_i}$ and $\mathcal{Z}$ are independent, $(c)$ holds by Lemma~\ref{Lemma:Swap} with $W\leftarrow (U,F_{z_{[i-1]}})$, $F_z\leftarrow F_{z_i}$, $\mathcal{S}\leftarrow \mathcal{S}$, and $\mathcal{S}'\leftarrow \ell_i'$, and because $\lim_{n\to \infty}I(F_{z_i},U,F_{z_{[i-1]}};\mathcal{Z})=0$ by~\eqref{Eq: C2}, $(d)$ holds because conditioning reduces entropy,  $(e)$ holds because $\hat F_{z_{[i-1]}}\triangleq (\hat{F}_{z_{j}})_{j \in [i-1]}$ is a function of $(Y^n,R_0,U)$, $(f)$ holds by Fano's inequality and \eqref{Eq: C1}, and $(g)$ holds by \eqref{Eq: C3}. Then,  we have 
\begin{align}
I(F_{\mathcal{Z}};U|\mathcal{Z})&=\sum_{\mathcal{S}\in \mathfrak{Z}}\mathbb{P}[\mathcal{Z}=\mathcal{S}]I(F_{\mathcal{Z}};U|\mathcal{Z}=\mathcal{S})\nonumber\\
&=o(n)\nonumber,
\end{align}
where the last equality holds by~\eqref{eq:laston}.
\end{proof}
Lemma~\ref{CL4} gives an upper bound on the entropy of the selected files by applying Lemmas~\ref{Lemma:XtYt} and~\ref{Lemma:MISwap}.

\begin{lemma} \label{CL4}
We have
\begin{align}
    H(F_{\mathcal{Z}})\leq \sum_{t=1}^nI(X_t;Y_t)+o(n).\nonumber
\end{align}
\end{lemma}
\begin{proof}
We have
 \begin{align}
            H(F_{\mathcal{Z}})
            &\stackrel{(a)}{=} H(F_{\mathcal{Z}}|\mathcal{Z})\nonumber \\
            &=H(F_{\mathcal{Z}}|U,\mathcal{Z})+I(F_{\mathcal{Z}};U|\mathcal{Z})\nonumber\\
            &\stackrel{(b)}{=}H(F_{\mathcal{Z}}|U,\mathcal{Z})+o(n)\nonumber\\
            &= I(F_{\mathcal{Z}};Y^n,R_0|U,\mathcal{Z})+H(F_{\mathcal{Z}}|Y^n,R_0,U,\mathcal{Z})+o(n)\nonumber \\
            &\stackrel{(c)}{=} I(F_{\mathcal{Z}};Y^n,R_0|U,\mathcal{Z})+H(F_{\mathcal{Z}}|\hat{F}_{\mathcal{Z}},Y^n,R_0,U,\mathcal{Z})+o(n)\nonumber \\
            &\stackrel{(d)}{=} I(F_{\mathcal{Z}};Y^n,R_0|U,\mathcal{Z})+ o(n) \nonumber \\ 
			&\stackrel{(e)}{\leq}\sum_{t=1}^{n}{I(X_t;Y_t )} +o(n), \nonumber
    \end{align}
where $(a)$ holds because $F_{\mathcal{Z}}$ and $\mathcal{Z}$ are independent, $(b)$ holds by Lemma~\ref{Lemma:MISwap}, $(c)$ holds because $ \hat{F}_{\mathcal{Z}}$ is a function of $(Y^n,R_0,U,\mathcal{Z})$, $(d)$ holds by Fano's inequality and \eqref{Eq: C1}, and $(e)$ holds by setting $t=n$ in Lemma~\ref{Lemma:XtYt}.
\end{proof}
 Then, we have 
\begin{align}
Kk
&=H(F_{\mathcal{Z}})\nonumber\\
&\leq \sum_{t=1}^{n} I(X_t;Y_t) + o(n)\nonumber \\  
&\leq n\max_{p_{X}} I(X;Y) + o(n), \label{Bound:MI}  
\end{align}
where the first inequality holds by Lemma~\ref{CL4}.
\subsection{Conditional entropy bound}\label{sec:CE-bound}
We now derive a second converse bound. We  start by deriving the following lemma.
\begin{lemma}\label{Lemma:addYn}
    For $\mathcal{S} \in \mathfrak{Z}$ and $z \in \mathcal{S}$, we have
    \begin{align}
        I(F_{z};Y^n,R_0|X^n,U,\mathcal{Z}=\mathcal{S})=0.\nonumber
    \end{align}
\end{lemma}
\begin{proof}
 For $t \in [n]$, we have 
 \begin{align}
     I(F_{[N]},R_1;Y^t,R_0|X^t, U_{[t]},\mathcal{Z})
     &\stackrel{(a)}{\le} I(F_{[N]},R_1;Y^t,R_0|X^t, U_{[t-1]},\mathcal{Z})\label{eq:withoutS}\\
     &=\!I(F_{[N]},R_1;Y^{t-1},R_0|X^t, U_{[t-1]},\mathcal{Z})\!+I(F_{[N]} ,R_1;Y_t|Y^{t-1},R_0,\mathcal{Z},X^t, U_{[t-1]})\nonumber\\
     &\stackrel{(b)}{=}I(F_{[N]},R_1;Y^{t-1},R_0|X^t, U_{[t-1]},\mathcal{Z})\nonumber \\
     &\le I(F_{[N]},R_1,X_t;Y^{t-1},R_0|X^{t-1}, U_{[t-1]},\mathcal{Z})\nonumber \\
     &=I(F_{[N]},R_1;Y^{t-1}\!\!\!,R_0|X^{t-1}\!\!, U_{[t-1]},\mathcal{Z})\!+\!I(X_t;Y^{t-1}\!\!,R_0|F_{[N]},R_1,X^{t-1}\!\!, U_{[t-1]},\mathcal{Z})\nonumber \\
     &\stackrel{(c)}{=}I(F_{[N]},R_1;Y^{t-1},R_0|X^{t-1}, U_{[t-1]},\mathcal{Z})\nonumber\\
     &\stackrel{(d)}{\le} I(F_{[N]},R_1;R_0|\mathcal{Z})\nonumber\\
     &=0,\nonumber
 \end{align}
 where $(a)$ holds by arguments similar to the proof of  Lemma~\ref{Lemma:Reduce_U_t}, $(b)$ holds because $(F_{[N]},R_1,Y^{t-1},R_0,\mathcal Z,U_{[t-1]},X^{t-1}) - X_t - Y_t$ forms a Markov chain, $(c)$ holds because $X_t$ is a function of $(F_{[N]},R_1,U_{[t-1]})$, and $(d)$ holds by repeating $t-1$ times the steps between $(a)$ and $(c)$. Then,  for $\mathcal{S} \in \mathfrak{Z}$ and $z \in \mathcal{S}$, we have
 \begin{align}
     I(F_{z};Y^n,R_0|X^n, U,\mathcal{Z}=\mathcal{S})\nonumber
     &\le I(F_{[N]},R_1;Y^n,R_0|X^n, U,\mathcal{Z}=\mathcal{S})\nonumber\\
     &\le \Big(\mathbb{P}[\mathcal{Z}=\mathcal{S}]\Big)^{-1} I(F_{[N]},R_1;Y^n,R_0|X^n, U,\mathcal{Z})\nonumber\\
     &=0,\nonumber
 \end{align}
 where the equality is satisfied by taking $t=n$ in~\eqref{eq:withoutS}.
\end{proof}

We now combine Lemmas~\ref{Lemma:Swap} and~\ref{Lemma:addYn} to show that, conditioned on $(X^n,U,\mathcal{Z})$, the  uncertainty about the unrequested files $F_{[N]\setminus \mathcal{Z}}$ is asymptotically negligible.
\begin{lemma}\label{Lemma:Swapentropy}
    We have 
    \begin{align}
        H(F_{[N]\setminus \mathcal{Z}}|X^n,U,\mathcal{Z})=o(n).\nonumber
    \end{align}
\end{lemma}
\begin{proof}
For $\mathcal{S}\in \mathfrak{Z}$ and $i\in [N-K]$, let $z_i\in [N]\setminus \mathcal{S}$, and $z\in \mathcal{S}$. Then, define $\mathcal{S}_i'\triangleq(\mathcal{S}\setminus\{z\})\cup\{z_i\}$. We have   \begin{align}
    H(F_{[N]\setminus \mathcal{Z}}|X^n,U,\mathcal{Z}=\mathcal{S})&\stackrel{(a)}{\le} \sum_{i=1}^{N-K}H(F_{z_i}|X^n,U,\mathcal{Z}=\mathcal{S})\nonumber\\
        &\stackrel{(b)}{=}\sum_{i=1}^{N-K}H(F_{z_i}|X^n,U,\mathcal{Z}=\mathcal{S}_i')+o(n)\nonumber\\
        &\stackrel{(c)}{\le} \sum_{i=1}^{N-K}H(F_{z_i}|Y^n,R_0,X^n,U,\mathcal{Z}=\mathcal{S}_i')+o(n)\nonumber\\
        &\stackrel{(d)}{\le}\sum_{i=1}^{N-K}H(F_{z_i}|Y^n,R_0,U,\mathcal{Z}=\mathcal{S}_i')+o(n)\nonumber\\
        &\stackrel{(e)}{=}o(n),\label{eq:laston2}
    \end{align}
    where $(a)$ holds by the chain rule and because conditioning reduces entropy, $(b)$ holds by Lemma~\ref{Lemma:Swap} with $W \leftarrow (X^n,U)$, $F_z\leftarrow F_{z_i}$, $\mathcal{S}\leftarrow\mathcal{S}_i'$, and $\mathcal{S}'\leftarrow \mathcal{S}$, and because $\lim_{n\to \infty}I( F_{z_i},X^n,U;\mathcal{Z})=0$ by~\eqref{Eq: C2}, $(c)$ holds by Lemma~\ref{Lemma:addYn},  $(d)$ holds because conditioning reduces entropy, and $(e)$ holds by Fano's inequality and \eqref{Eq: C1}.
     Then,  we have 
\begin{align}
H(F_{[N]\setminus \mathcal{Z}}|X^n,U,\mathcal{Z})\nonumber
&=\sum_{\mathcal{S}\in \mathfrak{Z}}\mathbb{P[\mathcal{Z}=\mathcal{S}]}H(F_{[N]\setminus \mathcal{Z}}|X^n,U,\mathcal{Z}=\mathcal{S})\nonumber\\
&=o(n),\nonumber
\end{align}
where the last equality holds by~\eqref{eq:laston2}.
\end{proof}
Lemma~\ref{Bound:entropy} gives an upper bound on the file size $k$ by applying Lemma~\ref{Lemma:Swapentropy}.
\begin{lemma}\label{Bound:entropy}
For any $K$-out-of-$N$ OT scheme, the file size $k$ satisfies 
\begin{align}
        (N-K)k\le H(X^n|Y^n)+o(n).\nonumber
    \end{align}
\end{lemma}
\begin{proof} We have
    \begin{align}
    (N-K)k&=H(F_{[N]\setminus \mathcal{Z}})\nonumber\nonumber\\
          &\stackrel{(a)}{=}H(F_{[N]\setminus \mathcal{Z}}|U,Y^n,R_0,\mathcal{Z})+o(n)\nonumber\\
          &\stackrel{(b)}{\le} H(F_{[N]\setminus \mathcal{Z}},X^n|U,Y^n,R_0,\mathcal{Z})+o(n)\nonumber\\
          &=\!H(X^n|U,Y^n,R_0,\mathcal{Z})\!+\!H(F_{[N]\setminus \mathcal{Z}}|X^n\!,U\!,Y^n\!,\!R_0,\!\mathcal{Z})\!+\!o(n)\nonumber\\
          &\stackrel{(c)}{\le} H(X^n|U,Y^n,R_0,\mathcal{Z})+H(F_{[N]\setminus \mathcal{Z}}|X^n,U,\mathcal{Z})+o(n)\nonumber\\
          &\stackrel{(d)}{=} H(X^n|U,Y^n,R_0,\mathcal{Z})+o(n)\nonumber\\
          &\stackrel{(e)}{\le} H(X^n|Y^n)+o(n),\nonumber
\end{align}
where $(a)$ holds by \eqref{Eq: C3}, $(b)$ holds by the chain rule, $(c)$ holds because conditioning reduces entropy, $(d)$ holds by Lemma~\ref{Lemma:Swapentropy}, and $(e)$ holds because conditioning reduces entropy.
\end{proof}
Then,  we have 
\begin{align}
 Kk
    &\stackrel{(a)}{\le}\frac{K}{N-K}H(X^n|Y^n)+o(n)\nonumber\\
    &\stackrel{(b)}{\leq}\frac{K}{N-K} \sum_{t=1}^{n} H(X_t|Y_t) + o(n)\nonumber \\ 
    &\stackrel{(c)}{=} n\frac{K}{N-K}H( X_T| Y_{T},T) + o(n) \nonumber \\ 
    &\le n\frac{K}{N-K}H( X_T| Y_{T}) + o(n) \nonumber \\ 
    &\leq n\frac{K}{N-K}\max_{p_{X}} H(X|Y) + o(n), \nonumber 
\end{align}
where $(a)$ holds by Lemma~\ref{Bound:entropy}, $(b)$ holds by the chain rule and because conditioning reduces entropy, and $(c)$ holds because $T$ is a uniform random variable over $[n]$ and independent of $(X^n,Y^n)$.

\section{Concluding Remarks}\label{sec:concluding}
In this paper, we characterized the capacity of $K$-out-of-$N$ OT over a BEC with a public noiseless channel under the honest-but-curious model, showing that
$C^{\binom{N}{K}}=\min\Big(\frac{K\epsilon}{N-K},\,1-\epsilon\Big).$ 
We studied a reduction-based achievability scheme that constructs a $1$-out-of-$N$ OT protocol from repeated uses of a $1$-out-of-$M$ OT primitive and  characterized the values of $M$ for which this reduction is optimal. 
Finally, for generalized erasure channels, we derived an achievable $K$-out-of-$N$ OT rate and showed its optimality when $\epsilon \ge \frac{N-K}{N}$.
\appendices

\section{\texorpdfstring{Achievability of Theorem~\ref{Theorem_2_def}}{Achievability of Theorem~2}}\label{sec:Achievability of Theorem 1}
 We begin with an example showing how   a $1$-out-of-$M$ scheme can be used to construct a $1$-out-of-$N$ scheme. We then present a general coding scheme to perform this reduction and its analysis in Appendices~\ref{CODS4} and~\ref{CODS42}, respectively.
\begin{example}\label{Ex:optimal_reduction}
Let $\epsilon=\frac34$ and $N=9$. One can reduce a $1$-out-of-$9$ OT problem to two uses of a $1$-out-of-$5$ OT protocol. Specifically, the server can group the files into two collections: $C_1=(F_1,F_2,F_3,F_4,Q_1)$ and
$C_2=(F_5\oplus Q_1,\;F_6\oplus Q_1,\;F_7\oplus Q_1,\;F_8\oplus Q_1,\;F_9\oplus Q_1),$
where $Q_1$ is an independent random string. If the client wants $F_7$, then in the first round it uses the $1$-out-of-$5$ primitive to obtain $Q_1$, and in the second round it uses the same primitive to obtain $F_7\oplus Q_1$. Combining the two outputs gives $F_7=(F_7\oplus Q_1)\oplus Q_1.$
Similarly, the client can recover any desired file, while every other file remains hidden by the security of the underlying $1$-out-of-$5$ OT instances. 
\end{example}
\subsection{Coding scheme} \label{CODS4}
For $M$ such that $2\le M<N$, the client and the server follow Algorithm~\ref{AG4} with  
\begin{align}\label{def:gammma}
    \gamma \triangleq \left\lceil \frac{N-1}{M-1}\right\rceil.
\end{align} 

\begin{algorithm*}
\caption{$1$-out-of-$N$ OT Scheme from $1$-out-of-$M$ OT Primitive} \label{AG4}
\begin{algorithmic}[1]
\setcounter{ALG@line}{0}
\Require  $\gamma-1$ sequences $(Q_t)_{t\in[\gamma-1]}$, each uniformly distributed over $\{0,1\}^k$, and define $Q_0 \triangleq 0^k$.
\Statex Define $L \triangleq N-(M-1)(\gamma-1)$. For $L<M$, let $(D_j)_{j\in[M-L]}$ be independent uniform sequences in $\{0,1\}^k$.
\State The server forms $(C_{t})_{t\in [\gamma]}$ as follows: $\forall t \in [\gamma-1]$\label{AG2DC4}, 
\begin{align} 
    (C_t[1], \ldots, C_t[M])&=(F_{(t-1)(M-1)+1}\oplus  Q_{t-1}, \ldots, F_{t(M-1)}\oplus Q_{t-1}, Q_t\oplus Q_{t-1}), \nonumber\\
    (C_{\gamma}[1], \ldots, C_{\gamma}[L]) &= (F_{(\gamma-1)(M-1)+1} \oplus Q_{\gamma-1}, \ldots, F_{N} \oplus Q_{\gamma-1}).\nonumber
\end{align}
Define $C_t \triangleq
\begin{cases}
    (C_{t}[1], \ldots, C_{t}[M]) & \text{if } t \in [\gamma-1],\\
    (C_{\gamma}[1], \ldots, C_{\gamma}[L],D_1,\ldots,D_{M-L}) & \text{if } t = \gamma.
\end{cases}$

\State  The client forms $(Z_t)_{t\in [\gamma] }$ as follows:
\begin{align}
   Z_t \triangleq 
\begin{cases}
    \Big((Z-1) \bmod (M -1)\Big)+1 & \text{if } t =\left\lceil \frac{Z}{M-1} \right\rceil, \nonumber\\
   M & \text{otherwise }
   \end{cases}.
\end{align}

\State  \textbf{for} $t \in [\gamma]$ \textbf{do}
\item   The server and client execute a $1$-out-of-$M$ OT primitive with server input the $M$ strings in $C_t$ and client choice $Z_t$. 
\State  \textbf{end for}
\State By Lines 3--5, the client can form \begin{align}
    F_{Z}=\bigoplus^{\min\Big(\left\lceil \frac{Z}{M-1} \right\rceil,\gamma\Big)}_{t=1}C_{t}[Z_t]. \nonumber
\end{align}
\end{algorithmic}
\end{algorithm*}
\subsection{Coding scheme analysis} \label{CODS42}
The security argument is analogous to that in the proof
of~\cite[Theorem~6]{10.1007/3-540-48910-X_4} and is therefore omitted.
 
To compute the achieved rate, we normalize the file length by $n\gamma$, because the channel is used $n\gamma$ times over the $\gamma$ rounds defined in~\eqref{def:gammma}. The achievable rate of the reduction is
\begin{align}
\frac{|F_Z|}{\gamma n}
&=\frac{1}{\gamma}\frac{|C_t[Z_t]|}{n}\nonumber\\
&\xrightarrow{n \to \infty} \frac{\min\left(\frac{\epsilon}{M-1},1-\epsilon\right)-\delta}{\left\lceil \frac{N-1}{M-1}\right\rceil}  \nonumber,
\end{align}
where  the limit follows from~\eqref{eq:AGK-BEC_ach_rate} applied to $K=1$ and $N=M$ messages.
\section{Proof of Theorem~\ref{Theorem:GEC_achievable_Rate}}\label{Sec:ach_for_GEC}
We begin with a secret-key result that will be used in the GEC achievability scheme.
\begin{proposition}\cite[Proposition~1]{Ahlswede2013}
\label{prop:SecretKeyGEC}
Let $(A,B)\in\mathcal A\times\mathcal B$ and $(\tilde A,T)\in\mathcal A\times\mathcal T$ be random pairs such that $A$ and $\tilde A$ have the same distribution. Let
$(A^n,B^n)\in\mathcal A^n\times\mathcal B^n$ and
$(\tilde A^n,T^n)\in\mathcal A^n\times\mathcal T^n$
consist of $n$ i.i.d.\ copies of $(A,B)$ and $(\tilde A,T)$,
respectively.
For any $\delta>0$ and sufficiently large $n$, there exist functions
$\zeta:\mathcal{A}^{n}\to\mathcal F_{n}$, 
$\kappa:\mathcal{A}^{n}\to\{0,1\}^{k}$, and $\hat\kappa:\mathcal F_n\times\mathcal B^n\to\{0,1\}^k$
such that, for some $\eta> 0$,
\begin{align}
k&=n\Bigl(I(A;B)-I(\tilde{A};T)-\delta\Bigr),\nonumber\\
\mathbb{P}\!\Big[\hat{\kappa}\Big(\zeta(A^n),B^n\Big)\neq \kappa(A^n)\Big]&\le 2^{-n\eta}, \nonumber\\
\lim_{n\to \infty}k-H\!\Bigl(\kappa(\tilde A^{n})| \zeta(\tilde A^{n}),T^{n}\Bigr)&= 0.\nonumber
\end{align}
   
\end{proposition}

\subsection{Coding scheme}\label{subsec: coding scheme GEC}

The participants follow Algorithm~\ref{AGK-GEC}, which has a similar structure to Algorithm~\ref{AGK-BEC}: the client identifies the non-erased and erased positions from the channel output and associates the requested files with the non-erased positions and the remaining files with the erased positions. In the BEC case, the server masks the files directly using the transmitted symbols at the selected positions. For the GEC, in contrast, the masking strings are constructed via Proposition~\ref{prop:SecretKeyGEC} rather than taken directly from the transmitted symbols.

\begin{algorithm}    
\caption{$K$-out-of-$N$ OT Scheme for the GEC} \label{AGK-GEC} 
\begin{algorithmic}[1]
 \setcounter{ALG@line}{0}
 \State Fix an input distribution $P_X$ on $\mathcal X$ and a constant $\delta>0$.
\State Consider a GEC $(\mathcal X,W_{Y|X},\mathcal Y)$ with parameter $0<\epsilon<1$ and erasure set $\mathcal{Y}_1$. Then, define $\mathcal Y_0\triangleq \mathcal Y\setminus\mathcal Y_1$, and write the GEC as \begin{align}
    W(y|x)=
\begin{cases}
(1-\epsilon)W_0(y|x), & y\in\mathcal Y_0,\\
\epsilon W_1(y), & y\in\mathcal Y_1.
\end{cases}\nonumber\end{align}
Define the joint distributions
\begin{align}
P_{XY}^{(0)}(x,y)
&\triangleq P_X(x)W_0(y|x),
& y\in\mathcal Y_0,\nonumber\\
P_{XY}^{(1)}(x,y)
&\triangleq P_X(x)W_1(y),
& y\in\mathcal Y_1.\nonumber
\end{align}
\State Let $(X^{(0)m},Y^{(0)m})$ consist of $m$ i.i.d.\ copies of
$(X_0,Y_0)\sim P_{XY}^{(0)}$, and let
$(\tilde{X}^{(1)m},Y^{(1)m})$ consist of $m$ i.i.d.\ copies of
$(\tilde{X}_1,Y_1)\sim P_{XY}^{(1)}$,
where $X_0$ and $\tilde{X}_1$ are identically distributed according to $P_X$ and we write $X^{(1)m}\triangleq\tilde X^{(1)m}$. 
Apply Proposition~\ref{prop:SecretKeyGEC} to 
$(X_0,Y_0)$ and $(\tilde X_1,Y_1)$
with block length $m$ and $\delta\leftarrow\frac{\delta}{1-\epsilon}$,
where $m$ is defined in~\eqref{eq:mdef}. Fix the resulting key length $k_{\mathrm{GEC}}$ and functions
$\zeta:\mathcal X^m\to\mathcal F_m$,
$\kappa:\mathcal X^m\to\{0,1\}^{k_{\mathrm{GEC}}}$, and
$\hat\kappa:\mathcal F_m\times\mathcal Y_0^m\to\{0,1\}^{k_{\mathrm{GEC}}}$. For $s\in[N]$, let the file be $F_s\in\{0,1\}^{k_{\mathrm{GEC}}}$.  \label{Line: AGK-GEC secret key}
\State The server sends an i.i.d.\ sequence $X^n \sim \prod_{t=1}^n P_X$  over the GEC. \label{Line: AGK-GEC Line 1}
\State Upon observing $Y^n$, the client proceeds as follows:\label{Line: AGK-GEC Line 2}
\begin{algsubstates}
  \State Define \label{Line: AGK-GEC Line 3} 
        $$\mathcal G \triangleq \{i\in[n]: Y_i\in\mathcal Y_0\}, \quad
        \mathcal B \triangleq \{i\in[n]: Y_i\in\mathcal Y_1\}. $$
    \State Follow Line~\ref{Line: AGK-BEC Line 4} of Algorithm~\ref{AGK-BEC}.
\State Follow Line~\ref{Line: AGK-BEC Line 5} of
Algorithm~\ref{AGK-BEC}.
\end{algsubstates}
\State Follow Line~\ref{Line AGK-BEC Nline 3} of Algorithm~\ref{AGK-BEC}.
If the client aborts, it informs the server over the public channel, and the protocol terminates. Otherwise, the client proceeds as
follows:\label{Line: AGK-GEC abort}
\begin{algsubstates}
\State  Follow Line~\ref{Line: AGK-BEC Line 6} of Algorithm~\ref{AGK-BEC}.
\State Follow Line~\ref{Line: AGK-BEC Line 7} of Algorithm~\ref{AGK-BEC} to construct the index sets $(\mathcal{L}_s)_{s\in[N]}$, which satisfy $\mathcal{L}_s\subseteq\mathcal{G}$ for $s\in\mathcal{Z}$ and $\mathcal L_s\subseteq\mathcal{B}$ for $s\in[N]\setminus\mathcal{Z}$.\label{Line: AGK-GEC construct sets}
\State Follow Line~\ref{Line: AGK-BEC Line 8} of Algorithm~\ref{AGK-BEC} to send $\mathcal{L}_{[N]}\triangleq(\mathcal{L}_s)_{s\in[N]}$ to the server over the public channel.\end{algsubstates}
\State For $s\in[N]$, define
$\zeta_s\triangleq\zeta(X^n[\mathcal L_s])$ and
$\kappa_s\triangleq\kappa(X^n[\mathcal L_s])$. The server sends
$(\zeta_s,F_s\oplus\kappa_s)_{s\in[N]}$ over the public channel.
\label{Line: AGK-GEC Line 9}
\State For each $s\in\mathcal Z$, since
$\mathcal L_s\subseteq\mathcal G$, every component of
$Y^n[\mathcal L_s]$ belongs to $\mathcal Y_0$.
The client computes
$\hat\kappa_s\triangleq
\hat\kappa\bigl(\zeta_s,Y^n[\mathcal L_s]\bigr)$
and then computes
$\hat F_s\triangleq
(F_s\oplus\kappa_s)\oplus\hat\kappa_s$.
\label{Line: AGK-GEC Line 10}
\end{algorithmic}
\end{algorithm}

\subsection{Reliability}\label{subsec: reliability GEC}
In this subsection, we prove~\eqref{Eq: C1}. By
Line~\ref{Line: AGK-GEC Line 10} of Algorithm~\ref{AGK-GEC}, the client
recovers a requested file $F_s$, $s\in\mathcal Z$, by removing the
one-time pad $\kappa_s$, which it must first reconstruct from its own
observation $Y^n[\mathcal L_s]$ and the public message $\zeta_s$.
Proposition~\ref{prop:SecretKeyGEC} guarantees that this reconstruction
succeeds with high probability, but only when the block
$(X^n[\mathcal L_s],Y^n[\mathcal L_s])$ is distributed as the source
$(X^{(0)m},Y^{(0)m})$ for which the code was designed in
Line~\ref{Line: AGK-GEC secret key} of Algorithm~\ref{AGK-GEC}. We therefore first determine the
distribution of these blocks in Lemma~\ref{lem:GEC blocks}. \eqref{Eq: C1} then follows because the
protocol aborts with vanishing probability.

Let $\mathcal A_n$ denote the abort event in
Line~\ref{Line: AGK-GEC abort} of Algorithm~\ref{AGK-GEC} and $\mathcal A_n^c$ its complement. Define
\begin{align}
\epsilon_i
\triangleq
\begin{cases}
1-\epsilon, & i=0,\\
\epsilon, & i=1,
\end{cases}
\label{eq:GEC-epsilon-i}
\end{align}
and define $W_1(y|x)\triangleq W_1(y).$
Then, for $i\in\{0,1\}$ and $y\in\mathcal Y_i$, we have
\begin{align}
W(y|x)&=\epsilon_iW_i(y|x),
\label{eq:GEC-channel-i}\\
P_{XY}^{(i)}(x,y)&=P_X(x)W_i(y|x).
\label{eq:GEC-PXY-i}
\end{align}

\begin{lemma}\label{lem:GEC blocks}
Fix $\mathcal{S}\in\mathfrak{Z}$, $s\in[N]$, and a realization $\ell_{[N]}\triangleq(\ell_i)_{i\in[N]}$ of $\mathcal L_{[N]}$
for which the event
$\mathcal{D}
\triangleq \{\mathcal Z=\mathcal S,\mathcal A_n^c,
\mathcal L_{[N]}=\ell_{[N]}\}$
has a positive probability.
Let
$i_s\triangleq
\begin{cases}
0, & s\in\mathcal S,\\
1, & s\notin\mathcal S,
\end{cases}$ and
$\mathcal{J}\triangleq[n]\setminus \ell_s
$.
Then, for every $x^m\in\mathcal X^m$ and
$y^m\in\mathcal Y_{i_s}^m$, we have
\begin{align}
\mathbb P\Big[
X^n[\ell_s]=x^m,
Y^n[\ell_s]=y^m\Big|\mathcal D\Big]=
\mathbb P\big[
X^{(i_s)m}=x^m,
Y^{(i_s)m}=y^m
\big].
\label{eq:GEC-block-law}
\end{align}
Moreover,  for every
$x^m\in\mathcal X^m$, $y^m\in\mathcal Y_{i_s}^m$,
$a^{n-m}\in\mathcal X^{n-m}$, and
$b^{n-m}\in\mathcal Y^{n-m}$, we have
\begin{align}
&\mathbb P\Big[
X^n[\ell_s]=x^m,Y^n[\ell_s]=y^m,
X^n[\mathcal J]=a^{n-m},Y^n[\mathcal J]=b^{n-m}
\Big|\mathcal D\Big]
\nonumber\\
&=
\mathbb P\Big[
X^n[\ell_s]=x^m,Y^n[\ell_s]=y^m
\Big|\mathcal D\Big]
\mathbb P\Big[
X^n[\mathcal J]=a^{n-m},Y^n[\mathcal J]=b^{n-m}
\Big|\mathcal D\Big].
\label{eq:GEC-block-indep}
\end{align}
\end{lemma}

\begin{proof}
For $t\in[n]$, let $E_t\triangleq1$ if $Y_t\in\mathcal Y_1$ and
$E_t\triangleq0$ otherwise, and let $E^n\triangleq(E_t)_{t\in[n]}$, so that
$\mathcal G=\{t:E_t=0\}$ and $\mathcal B=\{t:E_t=1\}$. Write
$\ell_s=\{t_1,\dots,t_m\}$ with $t_1<\dots<t_m$.

We first prove \eqref{eq:GEC-block-law}. For $A\triangleq
\{X^n[\ell_s]=x^m,Y^n[\ell_s]=y^m\}$, we have
\begin{align}
\mathbb P\Big[
A\Big|\,\mathcal D\Big]&\stackrel{(a)}{=}\sum_{e^n,r_0,z}
\mathbb P[A|
\mathcal D,E^n=e^n,R_0=r_0,\mathcal Z=z]
\mathbb P[E^n=e^n,R_0=r_0,\mathcal Z=z|\mathcal D]\nonumber\\
&=
\mathbb E\Big[
\mathbb P[A|\mathcal D,E^n,R_0,\mathcal Z]
\Big|\mathcal D
\Big]\nonumber\\
&\stackrel{(b)}{=}
\mathbb E\Big[
\mathbb P\big[A\,\big|\,E^n,R_0,\mathcal Z\big]
\,\Big|\,\mathcal D\Big]\nonumber\\
&\stackrel{(c)}{=}
\mathbb E\Big[
\mathbb P\big[A\,\big|\,E^n\big]
\,\Big|\,\mathcal D\Big]\nonumber\\
&\stackrel{(d)}{=}
\mathbb E\Big[
\prod_{j=1}^m\mathbb P\big[X_{t_j}=x_j,Y_{t_j}=y_j\,\big|\,E_{t_j}\big]
\,\Big|\,\mathcal D\Big]\nonumber\\
&\stackrel{(e)}{=}\mathbb E\Big[
\prod_{j=1}^m\mathbb P\big[X_{t_j}=x_j,Y_{t_j}=y_j\,\big|\,E_{t_j}=i_s\big]
\,\Big|\,\mathcal D\Big]\nonumber\\
&\stackrel{(f)}{=}\prod_{j=1}^m
\mathbb P\big[X_{t_j}=x_j,Y_{t_j}=y_j
\big|E_{t_j}=i_s\big]
\nonumber\\
&=
\prod_{j=1}^m
\frac{
\mathbb P\big[X_{t_j}=x_j,Y_{t_j}=y_j,E_{t_j}=i_s
\big]}{\mathbb P[E_{t_j}=i_s]}
\nonumber\\
&\stackrel{(g)}{=}
\prod_{j=1}^m
\frac{\mathbb P\big[X_{t_j}=x_j,Y_{t_j}=y_j
\big]}{\epsilon_{i_s}}
\nonumber\\
&\stackrel{(h)}{=}
\prod_{j=1}^m
\frac{P_X(x_j)W(y_j|x_j)}{\epsilon_{i_s}}
\nonumber\\
&\stackrel{(i)}{=}
\prod_{j=1}^m
P_{XY}^{(i_s)}(x_j,y_j)
\nonumber\\
&\stackrel{(j)}{=}
\mathbb P\big[
X^{(i_s)m}=x^m,Y^{(i_s)m}=y^m
\big],\nonumber
\end{align}
where $(a)$ holds by the law of total probability,
$(b)$ holds because, by Lines~\ref{Line: AGK-GEC Line 3}--~\ref{Line: AGK-GEC construct sets} of Algorithm~\ref{AGK-GEC}, $\mathcal D$ is determined by
$(E^n,R_0,\mathcal Z)$,
$(c)$ holds because $(R_0,\mathcal Z)$ is independent of $(X^n,Y^n,E^n)$,
$(d)$ holds because the triples $(X_t,Y_t,E_t)_{t\in[n]}$ are i.i.d.\ by
Line~\ref{Line: AGK-GEC Line 1} and since the GEC is memoryless,
$(e)$ holds because on $\mathcal D$, $\ell_s\subseteq\mathcal G$ if
$s\in\mathcal S$ and $\ell_s\subseteq\mathcal B$ otherwise, thus
$E_t=i_s$ for every $t\in\ell_s$, $(f)$ holds because the resulting product is constant on $\mathcal D$,
$(g)$ holds because $y_j\in\mathcal Y_{i_s}$ implies $E_{t_j}=i_s$, and by \eqref{eq:GEC-epsilon-i}, $\mathbb P[E_{t_j}=i_s]=\epsilon_{i_s}$,
$(h)$ holds by the definition of the memoryless channel,
$(i)$ holds by~\eqref{eq:GEC-channel-i} and~\eqref{eq:GEC-PXY-i} ,
and $(j)$ holds by Line~\ref{Line: AGK-GEC secret key} of Algorithm~\ref{AGK-GEC}.

We now prove \eqref{eq:GEC-block-indep}. Define
 $B\triangleq
\{X^n[\mathcal J]=a^{n-m},Y^n[\mathcal J]=b^{n-m}\}.$
Then, we have
\begin{align}
\mathbb P[A\cap B|\mathcal D]
&\stackrel{(a)}{=}
\mathbb E\big[
\mathbb P[A\cap B| E^n]
\big|\mathcal D
\big]\nonumber\\
&\stackrel{(b)}{=}
\mathbb E\big[
\mathbb P[A| E^n]\,
\mathbb P[B| E^n]
\big|\mathcal D
\big]\nonumber\\
&\stackrel{(c)}{=}
\mathbb P[A|\mathcal D]\,
\mathbb E\big[
\mathbb P[B| E^n]
\big|\mathcal D
\big]\nonumber\\
&\stackrel{(d)}{=}
\mathbb P[A|\mathcal D]\,
\mathbb P[B|\mathcal D],\nonumber
\end{align}
where $(a)$ holds by steps $(a)$--$(c)$ of the proof of
\eqref{eq:GEC-block-law}, applied to the event $A\cap B$,
$(b)$ holds because, conditional on $E^n$, the pairs
$(X_t,Y_t)_{t\in[n]}$ are independent across $t$, and $\ell_s$ and $\mathcal J$ are disjoint,
  $(c)$ holds by steps $(d)$–$(j)$ of the proof of \eqref{eq:GEC-block-law}, which show that, on $\mathcal D$,  $\mathbb P[A| E^n]=\prod_{j=1}^mP_{XY}^{(i_s)}(x_j,y_j)$,  and by \eqref{eq:GEC-block-law}, this product equals $\mathbb P[A|\mathcal D]$,  and $(d)$ holds by steps $(a)$--$(c)$ of the proof of \eqref{eq:GEC-block-law}, applied to the event $B$.

\end{proof}

By Proposition~\ref{prop:SecretKeyGEC}, for sufficiently large
$m$, the functions fixed in Line~\ref{Line: AGK-GEC secret key}
satisfy, for some $\eta>0$,
\begin{align}
\mathbb P\Big[
\hat\kappa\big(\zeta(X^{(0)m}),Y^{(0)m}\big)
\neq\kappa(X^{(0)m})
\Big]
\le 2^{-m\eta}.
\label{eq:prop-recovery}
\end{align}
Define the client's estimate arbitrarily on the abort event
$\mathcal A_n$. Then, we have
\begin{align}
\mathbb P\big[\hat F_{\mathcal Z}\neq F_{\mathcal Z}\big]
&\stackrel{(a)}{\le}
\mathbb P[\mathcal A_n]
+\max_{\mathcal S,\ell_{[N]}}\,
\mathbb P\big[\hat F_{\mathcal S}\neq F_{\mathcal S}\,\big|\,
\mathcal Z=\mathcal S,\mathcal A_n^c,
\mathcal L_{[N]}=\ell_{[N]}\big]\nonumber\\
&\stackrel{(b)}{\le}
\mathbb P[\mathcal A_n]
+\max_{\mathcal S,\ell_{[N]}}\,
\sum_{s\in\mathcal S}
\mathbb P\big[\hat\kappa_s\neq\kappa_s\,\big|\,
\mathcal Z=\mathcal S,\mathcal A_n^c,
\mathcal L_{[N]}=\ell_{[N]}\big]\nonumber\\
&\stackrel{(c)}{=}
\mathbb P[\mathcal A_n]
+K\,\mathbb P\Big[
\hat\kappa\big(\zeta(X^{(0)m}),Y^{(0)m}\big)
\neq\kappa(X^{(0)m})
\Big]\nonumber\\
&\stackrel{(d)}{\le}
\mathbb P[\mathcal A_n]+K\,2^{-m\eta}\nonumber\\
&{\xrightarrow{n\to\infty}}0,\nonumber
\end{align}
where the maxima are over $\mathcal S\in\mathfrak Z$ and realizations
$\ell_{[N]}$ of $\mathcal L_{[N]}$ such that
$\mathbb P[\mathcal Z=\mathcal S,\mathcal A_n^c,
\mathcal L_{[N]}=\ell_{[N]}]>0$,
$(a)$ holds by conditioning on the abort event and then averaging over
$(\mathcal Z,\mathcal L_{[N]})$ on $\mathcal A_n^c$,
$(b)$ holds by Line~\ref{Line: AGK-GEC Line 10} of
Algorithm~\ref{AGK-GEC}, since $\hat F_s=F_s$ if and only if
$\hat\kappa_s=\kappa_s$, and by the union bound over the $K$ files in
$\mathcal S$,
$(c)$ holds by Lemma~\ref{lem:GEC blocks} with $i_s=0$, since for each
$s\in\mathcal S$,
$\bigl(X^n[\mathcal L_s],Y^n[\mathcal L_s]\bigr)$ conditioned on
$\mathcal Z=\mathcal S$, $\mathcal A_n^c$, and
$\mathcal L_{[N]}=\ell_{[N]}$ has the same distribution as
$(X^{(0)m},Y^{(0)m})$,
$(d)$ holds by~\eqref{eq:prop-recovery},
and the limit holds because $m\to\infty$ for $0<\delta<Kr$, and
$\mathbb P[\mathcal A_n]\to0$ by the argument of
Lemma~\ref{Lemma:mzero}, since the labeling rule is the same as in
Line~\ref{Line: AGK-BEC Line 4} of Algorithm~\ref{AGK-BEC} and
$(\mathds{1}\{t\in\mathcal G\})_{t\in[n]}$ is i.i.d.\
Bernoulli$(1-\epsilon)$.

\subsection{Client privacy: Ensuring privacy of file selection}\label{subsec: client privacy GEC}   
Define $\zeta_{[N]}\triangleq (\zeta_s)_{s\in[N]}$ and $U_{\mathrm{GEC}}\triangleq \bigl(\zeta_{[N]},(F_s\oplus \kappa_s)_{s\in[N]},\mathcal L_{[N]}\bigr).$ Because $U_{\mathrm{GEC}}$ is a deterministic function of $(F_{[N]},X^n,\mathcal L_{[N]})$, the derivation leading to \eqref{eq:clientprivacyAGK-BEC} applies with $U_{\mathrm{GEC}}$ in place of $U$. Consequently, $I(F_{[N]},X^n,U_{\mathrm{GEC}};\mathcal Z)=0.$
\subsection{Server privacy: Ensuring privacy of unselected files}
To prove that the unselected files remain private, we first establish the next lemma to show that the binary key associated with any unrequested file remains asymptotically hidden from the client.  By Proposition~\ref{prop:SecretKeyGEC}, the key length $k_{\mathrm{GEC}}$ and the functions $\kappa$ and $\zeta$ fixed in Line~\ref{Line: AGK-GEC secret key} of Algorithm~\ref{AGK-GEC} satisfy
\begin{align}
    \lim_{m\to\infty}\Big(k_{\mathrm{GEC}}
-H\big(\kappa(\tilde X^{(1)m})\,\big|\,\zeta(\tilde X^{(1)m}),Y^{(1)m}\big)\Big)
&=0.\label{eq:prop-secrecy}
\end{align}

\begin{lemma}\label{PCAGEC}
Fix $\mathcal S\in\mathfrak Z$ and $s\in[N]\setminus\mathcal S$. Define $\mathcal T_s\triangleq[N]\setminus\{s\}$. Then, we have
\begin{align}
\lim_{n\to\infty}I\big(\kappa_s;Y^n,\zeta_{[N]},(F_i\oplus\kappa_i)_{i\in\mathcal T_s},F_{[N]},R_0\,\big|\,\mathcal Z=\mathcal S\big)=0.\nonumber
\end{align}
\end{lemma}

\begin{proof}
Let $\mathcal J\triangleq[n]\setminus\mathcal L_s$, $\kappa_{\mathcal T_s}\triangleq(\kappa_i)_{i\in\mathcal T_s}$, and $\zeta_{\mathcal T_s}\triangleq(\zeta_i)_{i\in\mathcal T_s}$. We have
\begin{align}
&I\big(\kappa_s;Y^n,\zeta_{[N]},(F_i\oplus\kappa_i)_{i\in\mathcal T_s},F_{[N]},R_0\,\big|\,\mathcal Z=\mathcal S\big)\nonumber\\
&\stackrel{(a)}{=}I\big(\kappa_s;Y^n,\zeta_{[N]},(F_i\oplus\kappa_i)_{i\in\mathcal T_s},F_{[N]},R_0,\mathcal L_{[N]}\,\big|\,\mathcal Z=\mathcal S\big)\nonumber\\
&\stackrel{(b)}{=}I\big(\kappa_s;Y^n,\zeta_{[N]},\kappa_{\mathcal T_s},F_{[N]},R_0,\mathcal L_{[N]}\,\big|\,\mathcal Z=\mathcal S\big)\nonumber\\
&=I\big(\kappa_s;Y^n,\zeta_{[N]},\kappa_{\mathcal T_s},F_{[N]},\mathcal L_{[N]}\,\big|\,\mathcal Z=\mathcal S\big)+I\big(\kappa_s;R_0\,\big|\,Y^n,\zeta_{[N]},\kappa_{\mathcal T_s},F_{[N]},\mathcal L_{[N]},\mathcal Z=\mathcal S\big)\nonumber\\
&\stackrel{(c)}{\le}I\big(\kappa_s;Y^n,\zeta_{[N]},\kappa_{\mathcal T_s},F_{[N]},\mathcal L_{[N]}\,\big|\,\mathcal Z=\mathcal S\big)+I\big(X^n,F_{[N]};R_0\,\big|\,Y^n,\mathcal L_{[N]},\mathcal Z=\mathcal S\big)\nonumber\\
&\stackrel{(d)}{=}I\big(\kappa_s;Y^n,\zeta_{[N]},\kappa_{\mathcal T_s},F_{[N]},\mathcal L_{[N]}\,\big|\,\mathcal Z=\mathcal S\big)\nonumber\\
&=I\big(\kappa_s;\mathcal L_{[N]}\,\big|\,\mathcal Z=\mathcal S\big)+I\big(\kappa_s;Y^n,\zeta_s,\zeta_{\mathcal T_s},\kappa_{\mathcal T_s},F_{[N]}\,\big|\,\mathcal L_{[N]},\mathcal Z=\mathcal S\big)\nonumber\\
&\stackrel{(e)}{=}I\big(\kappa_s;Y^n,\zeta_s,\zeta_{\mathcal T_s},\kappa_{\mathcal T_s},F_{[N]}\,\big|\,\mathcal L_{[N]},\mathcal Z=\mathcal S\big)\nonumber\\
&=I\big(\kappa_s;Y^n,\zeta_s\,\big|\,\mathcal L_{[N]},\mathcal Z=\mathcal S\big)+I\big(\kappa_s;\zeta_{\mathcal T_s},\kappa_{\mathcal T_s},F_{[N]}\,\big|\,Y^n,\zeta_s,\mathcal L_{[N]},\mathcal Z=\mathcal S\big)\nonumber\\
&\stackrel{(f)}{\le}I\big(\kappa_s;Y^n,\zeta_s\,\big|\,\mathcal L_{[N]},\mathcal Z=\mathcal S\big)+I\big(\kappa_s,\zeta_s,Y^n[\mathcal L_s];Y^n[\mathcal J],\zeta_{\mathcal T_s},\kappa_{\mathcal T_s},F_{[N]}\,\big|\,\mathcal L_{[N]},\mathcal Z=\mathcal S\big)\nonumber\\
&=I\big(\kappa_s;Y^n,\zeta_s\,\big|\,\mathcal L_{[N]},\mathcal Z=\mathcal S\big)+I\big(\kappa_s,\zeta_s,Y^n[\mathcal L_s];Y^n[\mathcal J],\zeta_{\mathcal T_s},\kappa_{\mathcal T_s}\,\big|\,\mathcal L_{[N]},\mathcal Z=\mathcal S\big)\nonumber\\
&\quad+I\big(\kappa_s,\zeta_s,Y^n[\mathcal L_s];F_{[N]}\,\big|\,Y^n[\mathcal J],\zeta_{\mathcal T_s},\kappa_{\mathcal T_s},\mathcal L_{[N]},\mathcal Z=\mathcal S\big)\nonumber\\
&\stackrel{(g)}{=}I\big(\kappa_s;Y^n,\zeta_s\,\big|\,\mathcal L_{[N]},\mathcal Z=\mathcal S\big)+I\big(\kappa_s,\zeta_s,Y^n[\mathcal L_s];F_{[N]}\,\big|\,Y^n[\mathcal J],\zeta_{\mathcal T_s},\kappa_{\mathcal T_s},\mathcal L_{[N]},\mathcal Z=\mathcal S\big)\nonumber\\
&\stackrel{(h)}{=}I\big(\kappa_s;Y^n,\zeta_s\,\big|\,\mathcal L_{[N]},\mathcal Z=\mathcal S\big)\nonumber\\
&=H\big(\kappa_s\,\big|\,\mathcal L_{[N]},\mathcal Z=\mathcal S\big)-H\big(\kappa_s\,\big|\,Y^n,\zeta_s,\mathcal L_{[N]},\mathcal Z=\mathcal S\big)\nonumber\\
&\stackrel{(i)}{\le}k_{\mathrm{GEC}}-H\big(\kappa_s\,\big|\,Y^n,\zeta_s,\mathcal L_{[N]},\mathcal Z=\mathcal S\big)\nonumber\\
&\stackrel{(j)}{=}k_{\mathrm{GEC}}-H\big(\kappa_s\,\big|\,Y^n[\mathcal L_s],\zeta_s,\mathcal L_{[N]},\mathcal Z=\mathcal S\big)\nonumber\\
&\stackrel{(k)}{=}k_{\mathrm{GEC}}-H\big(\kappa(X^{(1)m})\,\big|\,\zeta(X^{(1)m}),Y^{(1)m}\big)\nonumber\\
&\xrightarrow{n\to\infty}0,\nonumber
\end{align}
where $(a)$ holds because $\mathcal L_{[N]}$ is a function of $(Y^n,\mathcal Z,R_0)$ by Line~\ref{Line: AGK-GEC construct sets} of Algorithm~\ref{AGK-GEC}, 
$(b)$ holds because, given $F_{[N]\setminus\{s\}}$, $(F_i\oplus\kappa_i)_{i\in\mathcal T_s}$ and $\kappa_{\mathcal T_s}$ are functions of each other,
$(c)$ holds by the chain rule and because $(\kappa_{[N]},\zeta_{[N]})$ is a function of $(X^n,\mathcal L_{[N]})$, $(d)$ holds because $(X^n,Y^n,F_{[N]})$ is independent of $(R_0,\mathcal Z)$,
$(e)$ holds because for every realization $\ell_{[N]}$ of $\mathcal L_{[N]}$, $X^n[\ell_s]$ is distributed according to $\prod_{j=1}^mP_X$ given $\mathcal Z=\mathcal S$ and $\mathcal L_{[N]}=\ell_{[N]}$, and the distribution of $\kappa_s=\kappa(X^n[\ell_s])$ does not depend on $\ell_{[N]}$,
$(f)$ holds by the chain rule and the nonnegativity of mutual information,
$(g)$ holds by~\eqref{eq:GEC-block-indep}, since $(\kappa_s,\zeta_s,Y^n[\mathcal L_s])$ is a function of $(X^n[\mathcal L_s],Y^n[\mathcal L_s])$ and $(Y^n[\mathcal J],\zeta_{\mathcal T_s},\kappa_{\mathcal T_s})$ is a function of $(X^n[\mathcal J],Y^n[\mathcal J])$, as $\mathcal L_i\subseteq\mathcal J$ for $i\in\mathcal T_s$ by the pairwise disjointness of $(\mathcal L_i)_{i\in[N]}$,
$(h)$ holds because $F_{[N]}$ is independent  of $(X^n,Y^n,R_0)$ given $\mathcal Z=\mathcal S$ while the other variables in this term are functions of $(X^n,Y^n,R_0)$ given $\mathcal Z=\mathcal S$,
$(i)$ holds because $\kappa_s\in\{0,1\}^{k_{\mathrm{GEC}}}$,
$(j)$ holds by~\eqref{eq:GEC-block-indep}, as in $(g)$,
$(k)$ holds by~\eqref{eq:GEC-block-law} applied with $i_s=1$, since $s\notin\mathcal S$: for every realization $\ell_{[N]}$ of $\mathcal L_{[N]}$, $(X^n[\ell_s],Y^n[\ell_s])$ is distributed as $(X^{(1)m},Y^{(1)m})$ given $\mathcal Z=\mathcal S$, and $\mathcal L_{[N]}=\ell_{[N]}$, so that $(\kappa_s,\zeta_s,Y^n[\ell_s])$ is distributed as $(\kappa(X^{(1)m}),\zeta(X^{(1)m}),Y^{(1)m})$,
and the limit holds by~\eqref{eq:prop-secrecy}, since $m\to\infty$ as $n\to\infty$.
\end{proof}
Then, we  use Lemma~\ref{PCAGEC} and Lemma~\ref{Lemma 10} to conclude the proof of server privacy. We have
\begin{align}
I(F_{[N]\setminus \mathcal{Z}};U,Y^n,\mathcal{Z},R_0)
&=I(F_{[N]\setminus \mathcal{Z}};(F_{i}\oplus \kappa_i)_{i\in[N]},\zeta_{[N]},\mathcal{L}_{[N]},Y^n,\mathcal{Z},R_0)\nonumber\\
&\stackrel{(a)}{=}I(F_{[N]\setminus \mathcal{Z}};(F_{i}\oplus \kappa_i)_{i\in[N]},\zeta_{[N]},Y^n,\mathcal{Z},R_0)\nonumber\\
&=\sum_{\mathcal S\in\mathfrak Z}
\mathbb P[\mathcal Z=\mathcal S]\, I(F_{[N]\setminus \mathcal{Z}};(F_{i}\oplus \kappa_i)_{i\in[N]},\zeta_{[N]},Y^n,R_0|\mathcal{Z}=\mathcal{S})\nonumber\\
&\stackrel{(b)}{=}\sum_{\mathcal S\in\mathfrak Z}
\mathbb P[\mathcal Z=\mathcal S]\,\sum_{j\in[N]\setminus \mathcal{Z}} I(F_{j};(F_{i}\oplus \kappa_i)_{i\in[N]}, \zeta_{[N]},Y^n, R_0|F_{[j-1]\setminus \mathcal{Z}},\mathcal{Z}=\mathcal{S})\nonumber\\
&\leq \sum_{\mathcal S\in\mathfrak Z}
\mathbb P[\mathcal Z=\mathcal S]\,\sum_{j\in[N]\setminus \mathcal{Z}} I(F_{j};(F_{i}\oplus \kappa_i)_{i\in[N]},\zeta_{[N]},Y^n,R_0,F_{[N]\setminus\{j\}}|\mathcal{Z}=\mathcal{S})\nonumber\\
&\stackrel{(c)}{\leq} \sum_{\mathcal S\in\mathfrak Z}
\!\mathbb P[\mathcal Z\!=\!\mathcal S]\!\!\!\! \sum_{j\in[N]\setminus \mathcal{Z}}\!\!\! \Big(\! I(F_{j};\!(F_{i}\oplus \kappa_i)_{i\in[N]\setminus\{j\}},\zeta_{[N]},\!Y^n\!\!,R_0,F_{[N]\setminus\{j\}}|\mathcal{Z}\!=\!\mathcal{S})\!+\!\! \lambda_{n,j}\! +\! \delta_{n,j} \!\Big)\nonumber\\
&\stackrel{(d)}{=} \sum_{\mathcal S\in\mathfrak Z}
\mathbb P[\mathcal Z=\mathcal S]\, \sum_{j\in[N]\setminus \mathcal{Z}} (\lambda_{n,j} + \delta_{n,j})\nonumber\\
&\xrightarrow{n \to \infty}0, \nonumber
\end{align}
where $(a)$ holds because $\mathcal{L}_{[N]}$ is a deterministic function of $Y^n$, $\mathcal{Z}$, and $R_0$, $(b)$ holds by the chain rule, $(c)$ holds by applying Lemma~\ref{Lemma 10} under the conditional distribution
given $\mathcal Z=\mathcal S$, with $A \leftarrow F_j$,  $B \leftarrow B_j$, $C \leftarrow F_j$, and $D \leftarrow \kappa_j$, using the definitions $B_j\triangleq ((F_{i}\oplus \kappa_i)_{i\in[N]\setminus\{j\}},\zeta_{[N]},Y^n,R_0,F_{[N]\setminus\{j\}}) $,   $\lambda_{n,j} \triangleq I(\kappa_j; B_j, F_j|\mathcal{Z}=\mathcal{S})$, and $\delta_{n,j} \triangleq k_{\mathrm{GEC}} - H(\kappa_j)$, $(d)$ holds because files are uniformly distributed and independent of the channel output, the index sets, and $R_0$, and the limit holds because $\lambda_{n,j}$ vanishes as $n \to \infty$ according to Lemma~\ref{PCAGEC}, and $\delta_{n,j}$ vanishes as $n \to \infty$ by \eqref{eq:prop-secrecy}. 
\subsection{Achievable rate}
Let $(X^{(0)m},Y^{(0)m})$ and $(\tilde X^{(1)m},Y^{(1)m})$ be as in Line~\ref{Line: AGK-GEC secret key} of Algorithm~\ref{AGK-GEC}. By
Proposition~\ref{prop:SecretKeyGEC}, the key length and the functions
fixed in Line~\ref{Line: AGK-GEC secret key} of Algorithm~\ref{AGK-GEC} satisfy 
\begin{align}
k_{\mathrm{GEC}}
=m\left(I(X_0;Y_0)-I(\tilde X_1;Y_1)-\frac{\delta}{1-\epsilon}\right).
\label{K_GEC}
\end{align}
The rate is 
\begin{align}
\frac{|F_{\mathcal{Z}}|}{ n}&=\frac{
K|F_{1}
|}{ n}\nonumber\\
&\stackrel{(a)}{=}{\frac{K|\kappa_1|}{ n}}\nonumber\\
&\stackrel{(b)}{=}\frac{Kk_{\mathrm{GEC}}}{n}\nonumber\\
&\xrightarrow{n \to \infty}\frac{\min\Big(\frac{K\epsilon}{N-K},1-\epsilon\Big)-\delta}{1-\epsilon}\Big(I(X;Y)-\delta\Big),\nonumber
\end{align}
where  $(a)$ holds by  Line~\ref{Line: AGK-GEC Line 9} in  Algorithm~\ref{AGK-GEC}, $(b)$ holds by Line~\ref{Line: AGK-GEC secret key} in  Algorithm~\ref{AGK-GEC}, and the limit holds by~\eqref{K_GEC}  and~\eqref{eq:rdef}. 
\section{Proof of Corollary~\ref{Cor:GEC}}\label{Sec:ProofcorGEC}

We have
\begin{align}
C_{\mathrm{GEC}}^{\binom{N}{K}}
&\geq\frac{\max_{P_X} I(X;Y)}{1-\epsilon}\,
\min\!\Big(\frac{K\epsilon}{N-K},\,1-\epsilon\Big) \nonumber\\
&=
\begin{cases}
\frac{K\epsilon}{(N-K)(1-\epsilon)}\max_{P_X} I(X;Y), & \epsilon<\frac{N-K}{N},\\
\max_{P_X} I(X;Y), & \epsilon\geqslant\frac{N-K}{N},
\end{cases}
\label{eq:cor-lower}
\end{align}
where the inequality holds by \eqref{achievablerate_GEC}.
For $\epsilon\geqslant\frac{N-K}{N}$,  \eqref{capacityconversebound} provides the upper bound  $C_{\mathrm{GEC}}^{\binom{N}{K}}\leqslant\max_{P_X} I(X;Y)$, which, together with \eqref{eq:cor-lower}, establishes $C_{\mathrm{GEC}}^{\binom{N}{K}}=\max_{P_X} I(X;Y)$.

\section{Proof of Lemma~\ref{Lemma:Swap}}\label{proofLemma:Swap}\label{app:C}

To prove Lemma~\ref{Lemma:Swap}, we use the following entropy bound.
\begin{lemma}[{\cite[Lemma~3]{Ahlswede2013}}] \label{imrelemma}
Let $A$, $B$, and $C$ be random variables defined over the finite sets $\mathcal{A}$, $\mathcal{B}$, and $\mathcal{C}$, respectively. For any $c_1,c_2 \in \mathcal{C}$ with $\mathbb{P}[C = c_1]=p > 0$ and $\mathbb{P}[C=c_2]=q > 0$, we have 
		\begin{align}
			| H(A| B, C = c_1) - H(A| B, C = c_2)| \nonumber \leq 3\log_2 | \mathcal{A}|\sqrt{\frac{(p + q) \ln 2}{2pq} I(A,B ; C)}  + 1. \nonumber
		\end{align}
\end{lemma}

For any $\mathcal{S}, \mathcal{S}' \in \mathfrak{Z}$, we have
$p\triangleq\mathbb{P}[\mathcal{Z}=\mathcal{S}] = \frac{1}{\binom{N}{K}}$ and
$q\triangleq\mathbb{P}[\mathcal{Z}=\mathcal{S}'] = \frac{1}{\binom{N}{K}}$.
Let $F_z$ denote the $z$-th file, $z\in[N]$, and let $\mathcal{F}\triangleq\{0,1\}^k$.
Consider a random variable $W$ such that $\lim_{n\to\infty} I(F_z, W; \mathcal{Z}) = 0$
for every $z\in[N]$. Then, by Lemma~\ref{imrelemma} with $A\leftarrow F_z$,
$\mathcal{A}\leftarrow\mathcal{F}$, $B\leftarrow W$, $C\leftarrow\mathcal{Z}$,
$c_1\leftarrow\mathcal{S}$, and $c_2\leftarrow\mathcal{S}'$, we have
\begin{align}
\big|H(F_z| W,\mathcal{Z}=\mathcal{S}) -H(F_z| W,\mathcal{Z}=\mathcal{S}')\big| \nonumber
&\le 3\log_2|\mathcal{F}|\sqrt{\frac{(p+q)\ln 2}{2pq}I(F_z, W; \mathcal{Z})} + 1 \nonumber \\
&= 3k \sqrt{\binom{N}{K} \ln 2} \sqrt{I(F_z, W; \mathcal{Z})} + 1\nonumber\\
&=o(n),\nonumber
\end{align}
where the last equality holds because  \begin{align}
    \lim_{n\to \infty}\frac{3k\sqrt{\binom{N}{K} \ln 2} \sqrt{I(F_z,W;\mathcal{Z})}+1}{n}
    &=\frac{3R}{K}\sqrt{\binom{N}{K} \ln 2} \lim_{n\to \infty}\sqrt{I(F_z,W;\mathcal{Z})}\nonumber\\
    &=0.\nonumber
\end{align}

 \bibliographystyle{IEEEtran}
 \bibliography{refskon}
\end{document}